\documentclass[journal]{IEEEtran}

\usepackage{graphicx}
\usepackage{amsmath}
\usepackage{float}
\usepackage{caption}
\usepackage{subcaption}
\usepackage{array}
\usepackage{subcaption}
\usepackage{multirow}
\usepackage{amsfonts}
\usepackage{booktabs}
\usepackage{tcolorbox}
\usepackage{graphicx}
\usepackage{multicol}
\usepackage{multirow}
\usepackage{rotating}
\DeclareGraphicsExtensions{.pdf,.png}

\input{Qcircuit.tex}

\usepackage{epstopdf}

\begin{document}
%

\title{Everything You Always Wanted to Know About Quantum Circuits}
%
%
%

\author{Edgard Muñoz-Coreas, Himanshu Thapliyal
\thanks{E. Mu\~{n}oz-Coreas is with the Department of Electrical Engineering, University of North Texas, Denton, TX. 76207 USA }
\thanks{H. Thapliyal is with the Department of Electrical Engineering and Computer Science, University of Tennessee, Knoxville, TN. 37966 USA. The version posted is the submitted version accepted for publication and the citation of the final published article is \cite{munoz-coreas_everything_2022}.}
\thanks{
}
}
%
\markboth{Journal of \LaTeX\ Class Files,~Vol.~14, No.~8, August~2015}%
{Shell \MakeLowercase{\textit{et al.}}: Bare Demo of IEEEtran.cls for IEEE Journals}
%



\maketitle


\section{Introduction}

The development of circuits for quantum computing has been motivated by the proliferation of quantum algorithms which promise up to a superpolynomial factor speedup over classical equivalents.  The quantum algorithms developed have the potential to impact fields such as number theory,  encryption, scientific computation \cite{Shor1997GeneralNonImageQuantumAlgorithm}  \cite{Cheung}  \cite{Montanaro}  \cite{Seroussi}   \cite{VanDam2008exponential} \cite{LeGall2017TriangleFinding} \cite{Bowregard} \cite{Aditya2018QuantumImageProcessing} \cite{HuWenWen2019QuantumWatermarking} \cite{ChengZhenwen2018QuantumWatermarking} \cite{Bhaskar2016QuantumScientificComputation}  \cite{Hallgren2007GeneralNonImageQuantumAlgorithm} \cite{Bravyi2011QuantumImageAlgorithm} \cite{Jordan2005science} \cite{Trout2015QuantumScientificCompiletion}.  The design of quantum algorithms remains an active area of research and new algorithms continue to appear in the literature (see \cite{Zoo} for a representative listing of quantum algorithms).  

In order to realize the potential performance gains of these proposed quantum algorithms, they must be implemented on quantum hardware.  The quantum computers developed by entities such as IBM or Honeywell are examples of quantum hardware platforms which could be used to implement quantum algorithms \cite{IBM2017QuantumNISQMachine} \cite{Hannes2017iontrapNISQ} \cite{IonQ2020QuantumNISQMachine} \cite{Honeywell2020NISQmachine} \cite{Edgard2019wow}.  To implement quantum algorithms on these hardware platforms, we require quantum datapath systems which are composed of quantum circuits.    

\textit{In this work we shall introduce the design and resource cost assessment of quantum circuits.}  These quantum circuits are composed of quantum gate networks.  Quantum machines developed from entities such as IBM and Honeywell support gate based quantum computation.  Gate based quantum circuit design has applications in fault tolerant quantum computation and in quantum circuit design automation \cite{Maslov} \cite{Devitt} \cite{Paler_2017FaultTolerant} \cite{Kiteav2005faulttolerant} \cite{Fowler2012QuantumSurfaceCodes} \cite{Steane1997SteaneCodes} \cite{Trout2015QuantumScientificCompiletion} \cite{LaoLingling2021QuantumCircuitCOmpilation}.  Each quantum gate represents a quantum mechanical operation.  As a result, the designer working with quantum circuits shall have to contend with novel properties and challenges.  For instance, quantum circuits are one-to-one and all information is preserved.   

The design of quantum circuits for the implementation of quantum algorithms has caught the attention of researchers.  Circuits for elementary functions such as basic arithmetic functions (such as addition or division) have been proposed such as \cite{Jayashree} \cite{Edgard2019multiplication}\cite{Kotiyal2014multiplier} \cite{Edgard2021addition} \cite{Saeedi2} \cite{Lidia} \cite{Edgard2019divider} \cite{Thapliyal2016addsub}.  These basic circuits are used as building blocks for more complex datapath systems such as higher level mathematical functions for applications in scientific computing, image processing or machine learning \cite{RiGuoZhou2017ImageInterpolation} \cite{FeiYan2017QuantumRotation} \cite{Edgard2018bilinear} \cite{Bhaskar2016QuantumScientificComputation} \cite{Neto2020quantumNeuralNet} \cite{Edgard2018JETCsqrt} \cite{Zhang2020quantumMatrixOperations}.  These functions in turn perform the computations called for in quantum algorithms.  As  a result, dedicated libraries of basic quantum arithmetic functions are included in quantum programming languages such as Quipper \cite{Quipper} and $LIQUi\vert\rangle$ \cite{LIQUi} and researchers continue to invest efforts in the design of increasingly more resource efficient quantum circuits.  \textit{In this work, we illustrate how elementary quantum circuits could be used to implement a quantum circuit for the image rotation operation.  The image rotation operation is a higher level mathematical function for that can be used for quantum realizations of image processing applications such as those shown in \cite{Iliyasu2013useofrotation} \cite{Lehmann1999useofrotation}  \cite{Haponen2003useofimagerotation} \cite{FeiYan2017QuantumRotation}.}    

In this work, we provide an overview of circuits for quantum computing.  In Section \ref{Overview-Section}, we introduce gates used in quantum computation then present resource cost measurements used to evaluate circuits made from these gates shown.  We then illustrate how the gates shown are then combined into quantum circuits for basic arithmetic functions in Section \ref{Quantum-circuit-arithmetic}.  In Section \ref{quantum-costs}, we demonstrate how to calculate the resource costs of quantum circuits.  We conclude this overview with Section \ref{example-circuit-implementation} by illustrating an application of the elementary quantum circuits shown in Section \ref{Quantum-circuit-arithmetic}.

\section{Overview Of Quantum Circuits}
\label{Overview-Section}

\begin{figure}[tbhp]
	
	\begin{tcolorbox}[title = Unique Quantum Computer Properties,fonttitle=\bfseries , colframe=blue!90!black, colback=blue!10!white]
		\begin{itemize}
			\item \textbf{Superposition:} 
			
			\includegraphics[width = 2.5in]{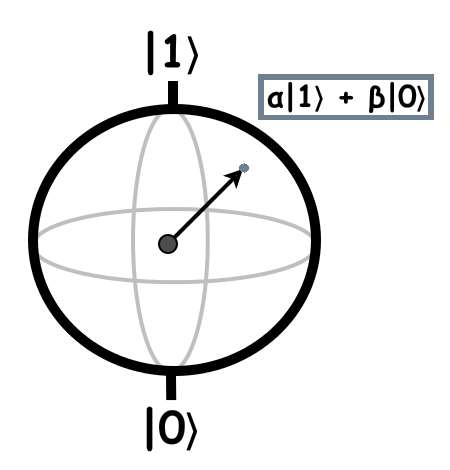}
			
			A qubit is probabilistic in nature. Thus, a qubit is $1$ with a probability $\alpha$ and $0$ with a probability $\beta$.  Thus, a qubit can assume an infinite range of values (any point on the surface of the sphere shown in the illustration above).  With gates we can adjust the probability values of a qubit.
			\item \textbf{Entanglement:} Two or more qubits can become interdependent.  Thus, information about one of the entangled qubits can allow us to glean information on the others.
			\item \textbf{Unitarity:}  For given quantum operation $U$, there exists a $U^\dagger$ which undoes the computation.  As a result, quantum computation is reversible in nature.  			
			\item \textbf{Measurement:}  The result of computation of a quantum circuit is read via measurement.  Consequently, qubits in superposition are either set to values $\ket{0}$ and $\ket{1}$.  Which of these computational basis states (or eigenstates) appears on the measured qubit shall depend on the associated state probabilities.

		\end{itemize}
		\caption{ Properties of quantum computer.  For more details, the reader is encouraged to see \cite{Chaung2011QuantumBook} \cite{sakurai_napolitano_2017} \cite{Edgard2019wow}. }
		\label{definitions-Calout}
	\end{tcolorbox}
	
\end{figure}

Quantum circuits based on gates represent sequences of controlled quantum mechanical operations.  These operations act on information stored in one or more quantum bits or qubits.  The physical implementation of the qubit shall depend on the underlying technology.   For instance, qubits are realized with Josephson junctions in superconducting quantum machines or qubits are individual ions in the case of a trapped ion quantum machine \cite{Edgard2019wow} \cite{Rigetti2019QuantumNISQMachine} \cite{IBM2017QuantumNISQMachine} \cite{Honeywell2020NISQmachine}.  Implementation differences aside, the quantum machine allows us to utilize the properties of quantum mechanics to perform useful work.  Such useful properties include superposition and entanglement.  Superposition is a consequence of the probabilistic nature of quantum machines and entanglement allows us to create dependencies between two or more qubits.  Figure \ref{definitions-Calout} introduces these properties. Thanks to superposition and entanglement it is possible for an $n$ qubit quantum circuit to operate on $2^n$ values simultaneously.  This is because, pursuant to superposition, we are operating not on fixed binary values but on probability distributions (see Figure \ref{definitions-Calout}).  As a result every value is present with an associated probability $p$.  The ability to make codependent qubits through entanglement is a powerful tool that allows us to, among other tasks, create fault tolerant quantum gate implementations \cite{Zhou2000FaultTolerantQuantumComputation} \cite{Devitt}.  

A consequence of working with quantum mechanical operations (or gates) is that they are unitary.  The unitary principal means that for a given quantum operation $U$, there exists a $U^\dagger$ such that the following expression is true:

\begin{tcolorbox}[fonttitle=\bfseries , colframe=green!90!black, colback=green!10!white]
	\begin{equation}
	U U^{\dagger} = I
	\label{encyclopedia-ee-eq6}
	\end{equation}
\end{tcolorbox}

\noindent
where $U$ is a quantum mechanical operation and $U^{\dagger}$ is the complex conjugate transpose of $U$.  

Thus, there exists an operation which undoes the work accomplished by a previous quantum mechanical operation.  Or, more specifically, for a given network of gates, there exists a gate network which shall undo the results of computation.  This means that \textit{quantum gates and quantum circuits are reversible in nature}.  As a result (i) information is not destroyed and (ii) the mapping between circuit inputs and outputs is one-to-one.  As a result the quantum circuit designer works exclusively with reversible gates and shall have to take into account new measures when evaluating the resource cost of a quantum circuit.  Section \ref{encycpedia-gate-section} introduces many of these gates and Section \ref{encycpedia-cost-measure-section} introduces the resource cost measures.          

To determine the result of computation, a contents of a quantum register needs to be measured.  Measurement is also used when the result of computation of a quantum circuit needs to be provided as input to a classical circuit.  To illustrate measurement we need a way to present quantum states.  We can represent a quantum state using vectors and matrices.  Consider a qubit (a two level quantum state) $X$.  The observable states of a qubit are $0$ and $1$ which we can represent in vector form as:  

\begin{tcolorbox}[fonttitle=\bfseries , colframe=green!90!black, colback=green!10!white]
\begin{equation}
	\ket{0} = \begin{bmatrix}
	1 \\ 
	0
	\end{bmatrix} \qquad \ket{1} = \begin{bmatrix}
	0 \\ 
	1
	\end{bmatrix} 
		\label{encyclopedia-ee-eq14}
\end{equation}
\end{tcolorbox}

So we can represent $X$ when in a superposition as follows:

\begin{tcolorbox}[fonttitle=\bfseries , colframe=green!90!black, colback=green!10!white]
\begin{equation}
	\ket{X} = \alpha \cdot \begin{bmatrix}
	1 \\ 
	0
	\end{bmatrix}  + \beta \cdot \begin{bmatrix}
	0 \\ 
	1
	\end{bmatrix} \equiv \begin{bmatrix}
	\alpha \\
	\beta 
	\end{bmatrix}
	\label{encyclopedia-ee-eq15}
\end{equation}
\end{tcolorbox}

The symbol around $X$ which appears in equation \ref{encyclopedia-ee-eq15} is a ``ket''.  It is a shorthand notation for the vectors presented in the equations.  If we were to take a complex conjugate of any of these vectors we can represent it using a ``bra''.  So $X^\dagger$ can be presented with a ``bra'' as follows: $ \langle X |$.  Collectively, this notation is called ``bra-ket'' notation and is widely used when describing quantum computing and the underlying physics.  

So now that we know how one can present mathematically quantum states such as qubits, we now turn our attention to measurement.  The quantum measurement task represents a projective quantum mechanical operation \cite{sakurai_napolitano_2017} \cite{Griffiths2004Introduction}.  In a projective operation, a qubit in superposition shall be set to one of a set of observable states (also called eigenstates) at the end of computation \cite{sakurai_napolitano_2017} \cite{Griffiths2004Introduction}. So, for the case of a qubit in superposition such as $\ket{X} = \alpha \cdot \ket{0} + \beta \cdot \ket{1}$, the observable states are $\ket{0}$ and $\ket{1}$.  

So, at the end of computation $\ket{X}$ shall be projected to $\ket{0}$ or $\ket{1}$.  The likelihood of seeing $\ket{0}$ and $\ket{1}$ shall depend on the associated probabilities $\alpha$ and $\beta$.  More generally given a quantum state $\ket{A}$ given as:

\begin{tcolorbox}[fonttitle=\bfseries , colframe=green!90!black, colback=green!10!white]
	\begin{equation}
	\ket{A} = \sum_{M} c_i \ket{m_i} \equiv \sum_{M} \ket{m_i} \langle m_i | A \rangle
	\label{encyclopedia-ee-eq5}	
	\end{equation}
\end{tcolorbox}

\noindent
where $m_i \in M$ is the $i$th observable state and $c_i$ is the associated probability.  In measurement, at the end of computation, $\ket{A}$ shall be projected to the observable $m_i$ (where $m_i \in M$) that enjoys the largest probability $c_i$.

\subsection{Quantum Gates}
\label{encycpedia-gate-section}

\begin{figure}[hbtp] 
	\centering
	\begin{tcolorbox}[title = How to Read a Quantum Circuit Diagram,fonttitle=\bfseries , colframe=blue!90!black, colback=blue!10!white]
		\includegraphics[width = 3.1in]{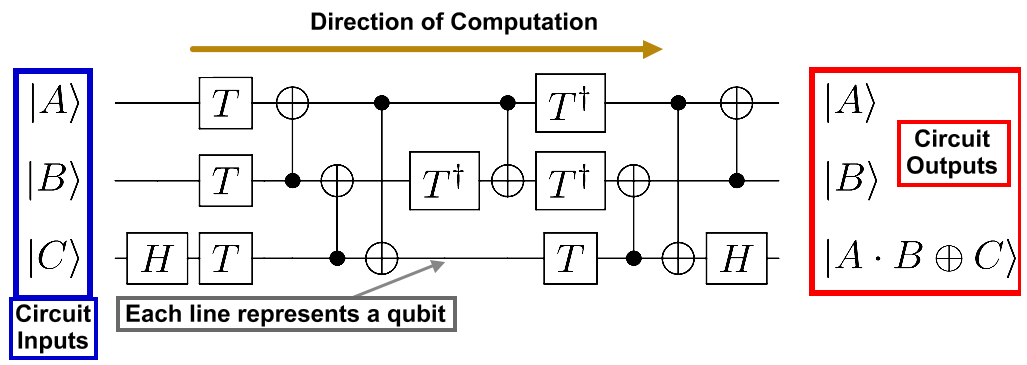}
		\caption{This marked up example of a quantum circuit diagram illustrates how to read a quantum circuit diagram.}
		\label{How-to-read-circuit}
		
	\end{tcolorbox}
\end{figure}

\begin{figure}[hbt]
	\centering
	
	\begin{tcolorbox}[title = Clifford+T Quantum Gates,fonttitle=\bfseries , colframe=blue!90!black, colback=blue!10!white]
		\begin{subfigure}[hb]{1in}
			Hadamard Gate
		\end{subfigure} \qquad \begin{subfigure}[hb]{.5in}
			\[
			\Qcircuit @C=0.7em @R=0.5em @!R{
				& \gate{H} &  \qw & &   }
			\]
		\end{subfigure} \qquad \begin{subfigure}[hb]{.75in}
			\centering
			$ \frac{1}{\sqrt{2}}
			\begin{bmatrix}
			1 & 1  \\
			1 & -1 
			\end{bmatrix} $
		\end{subfigure} \\ \begin{subfigure}[hb]{1in}
			
		\end{subfigure} \qquad \begin{subfigure}[hb]{.5in}
			
		\end{subfigure} \qquad \begin{subfigure}[hb]{.75in}
			
		\end{subfigure} \\ \begin{subfigure}[hb]{1in}
			T Gate
		\end{subfigure} \qquad \begin{subfigure}[hb]{.5in}
			\[
			\Qcircuit @C=0.7em @R=0.5em @!R{
				& \gate{T} &  \qw & &   }
			\]
		\end{subfigure} \qquad \begin{subfigure}[hb]{.75in}
			\centering
			$\begin{bmatrix}
			1 & 0  \\
			0 & e^{i \cdot \frac{\pi}{4}} 
			\end{bmatrix} $
		\end{subfigure} \\ \begin{subfigure}[hb]{1in}
			
		\end{subfigure} \qquad \begin{subfigure}[hb]{.5in}
			
		\end{subfigure} \qquad \begin{subfigure}[hb]{.75in}
			
		\end{subfigure} \\ \begin{subfigure}[hb]{1in}
			\flushleft
			Hermitian of T Gate
		\end{subfigure} \qquad \begin{subfigure}[hb]{.5in}
			\[
			\Qcircuit @C=0.7em @R=0.5em @!R{
				& \gate{T^{\dag}} &  \qw & &   }
			\]
		\end{subfigure} \qquad \begin{subfigure}[hb]{.75in}
			\centering
			$\begin{bmatrix}
			1 & 0  \\
			0 & e^{-i \cdot \frac{\pi}{4}} 
			\end{bmatrix} $
		\end{subfigure} \\ \begin{subfigure}[hb]{1in}
			
		\end{subfigure} \qquad \begin{subfigure}[hb]{.5in}
			
		\end{subfigure} \qquad \begin{subfigure}[hb]{.75in}
			
		\end{subfigure} \\ \begin{subfigure}[hb]{1in}
			Phase Gate
		\end{subfigure} \qquad \begin{subfigure}[hb]{.5in}
			\[
			\Qcircuit @C=0.7em @R=0.5em @!R{
				& \gate{S} &  \qw & &   }
			\]
		\end{subfigure} \qquad \begin{subfigure}[hb]{.75in}
			\centering
			$\begin{bmatrix}
			1 & 0  \\
			0 & i 
			\end{bmatrix} $
		\end{subfigure} \\ \begin{subfigure}[hb]{1in}
			
		\end{subfigure} \qquad \begin{subfigure}[hb]{.5in}
			
		\end{subfigure} \qquad \begin{subfigure}[hb]{.75in}
			
		\end{subfigure} \\ \begin{subfigure}[hb]{1in}
			\flushleft
			Hermitian of Phase Gate
		\end{subfigure} \qquad \begin{subfigure}[hb]{.5in}
			\[
			\Qcircuit @C=0.7em @R=0.5em @!R{
				& \gate{S^{\dag}} &  \qw & &   }
			\]
		\end{subfigure} \qquad \begin{subfigure}[hb]{.75in}
			\centering
			$\begin{bmatrix}
			1 & 0  \\
			0 & -i 
			\end{bmatrix} $
		\end{subfigure} \\ \begin{subfigure}[hb]{1in}
			
		\end{subfigure} \qquad \begin{subfigure}[hb]{.5in}
			
		\end{subfigure} \qquad \begin{subfigure}[hb]{.75in}
			
		\end{subfigure} \\ \begin{subfigure}[hb]{1in}
			NOT Gate
		\end{subfigure} \qquad \begin{subfigure}[hb]{.5in}
			\[
			\Qcircuit @C=0.7em @R=0.5em @!R{
				& \targ &  \qw & &   }
			\]
		\end{subfigure} \qquad \begin{subfigure}[hb]{.75in}
			\centering
			$\begin{bmatrix}
			0 & 1  \\
			1 & 0 
			\end{bmatrix}$
		\end{subfigure} \\ \begin{subfigure}[hb]{1in}
			
		\end{subfigure} \qquad \begin{subfigure}[hb]{.5in}
			
		\end{subfigure} \qquad \begin{subfigure}[hb]{.75in}
			
		\end{subfigure} \\ \begin{subfigure}[hb]{1in}
			Feynman (CNOT) Gate
		\end{subfigure} \qquad \begin{subfigure}[hb]{.5in}
			\[
			\Qcircuit @C=0.7em @R=0.5em @!R{
				& \ctrl{1} &  \qw & & \\ 
				& \targ &  \qw & & }
			\]
		\end{subfigure} \qquad \begin{subfigure}[hb]{.75in}
			\centering
			$\begin{bmatrix}
			1 & 0 & 0 & 0 \\
			0 & 1 & 0 & 0\\
			0 & 0 & 0 & 1 \\
			0 & 0 & 1 & 0 \\
			\end{bmatrix} $ 
		\end{subfigure}
		
		\caption{The quantum gate set used in this work.}
		\label{Clifford table}
		
	\end{tcolorbox}
\end{figure}

\begin{figure}

\begin{tcolorbox}[title = Controlled Gate Example,fonttitle=\bfseries , colframe=blue!90!black, colback=blue!10!white]

\centering

\begin{subfigure}[hbt]{1.35in}
 \[
			\Qcircuit @C=0.4em @R=0.5em @!R{
				\lstick{\ket{A}} & \ctrl{1} &  \qw & \rstick{\ket{A}} \\ 
				\lstick{\ket{B}} & \targ &  \qw & \rstick{\ket{A \oplus B}} }
			\]
		\end{subfigure} \qquad \begin{subfigure}[hbt]{1.35in}
		\[ \ket{B} = \begin{cases} 
			\ket{B} \text{ if } \ket{A} = 0 \\
		\ket{\overline{B}} \text{ if } \ket{A} = 1 \end{cases}
		\]
		\end{subfigure}
\caption{The CNOT gate is an example of a controlled gate.  Schematic diagram and mathematical description of operation are shown.}
\label{Controll-gate-Fig}
\end{tcolorbox}
\end{figure}	

We represent quantum gates either as (i) unitary matrices or (ii) schematic diagrams.  Figure \ref{Clifford table} presents the matrix and gate image representations for a frequently used quantum gate set called the Clifford+T gate set.  We shall focus on the Clifford+T gate set because it is (i) an approximately universal set and (ii) these can be made fault tolerant with existing quantum error correcting codes \cite{Fowler2012QuantumSurfaceCodes} \cite{Kiteav2005faulttolerant} \cite{Haah2017magicstate} \cite{Kliuchnikov2013approximationofCliffT} \cite{Miller} \cite{Devitt}.

Gates such as the Hadamard gate or T gate are examples of gates which produce a superposition state at the end of computation.  Access to gates with produce superpositions enables the quantum circuit designer to execute a richer set of possible computations.  For example, with a quantum computer, one can directly calculate the quantum Fourier transform shown below:

\begin{tcolorbox}[fonttitle=\bfseries , colframe=green!90!black, colback=green!10!white]
	\begin{equation}
	\ket{x} = \frac{1}{\sqrt{N}} \sum_{y = 0}^{N-1} e^\frac{2 \cdot \pi \cdot j \cdot x \cdot y}{N} \ket{y}
	\label{encyclopedia-ee-eq7}
	\end{equation}
\end{tcolorbox}

Equation \ref{encyclopedia-ee-eq7} is analogous to the discrete Fourier transform and is used in the implementation of algorithms such as those shown for integer factoring (see \cite{Shor1997GeneralNonImageQuantumAlgorithm}), the hidden linear structure problem (\cite{Niel2000QFTalgorithm})  or link invariant problems (see \cite{Krovi2015QFTalgotithm}).

 Gates such as the Hadamard and T gate are called one qubit gates and the CNOT gate is an example of a two qubit gate.  The CNOT gate (or Controlled NOT gate) is a two input operation where one input is referred to as the control qubit (qubit $\ket{A}$ in Figure \ref{Controll-gate-Fig}) and the second input is the target qubit (qubit $\ket{B}$ in Figure \ref{Controll-gate-Fig}).  As shown in Figure \ref{Controll-gate-Fig}, the value of the control qubit shall determine the result of computation seen on the target qubit.  Referring to the Figure, when $\ket{A = 1}$, $\ket{B}$ shall have the value $ 1 \oplus B \equiv \overline{B}$.  When $\ket{A = 0}$, $\ket{B}$ shall be unchanged at the end of computation.  The CNOT gate is an example of a controlled gate because the action of the gate (the NOT operation) is controlled (in this case by the value of qubit $\ket{A}$.  The CNOT gate is also referred to as a Feynman gate in the literature.

Figure \ref{How-to-read-circuit} shows how to read the quantum circuits shown in this work.  Quantum circuits are read left to right.  Each "line" shown in Figure \ref{How-to-read-circuit} represents a qubit.  Quantum gates are laid out as needed on the diagram to illustrate the computation tasks to be performed.  The quantum circuit diagram provides a temporal representation of the computations taking place on quantum machines.  Thus, in a quantum circuit diagram (such as the one shown in Figure \ref{How-to-read-circuit}) one is seeing how quantum mechanical operations are applied as a function of time to complete a given operation.

\begin{figure}
	
	\begin{tcolorbox}[title = The Toffoli and the Fredkin Gate ,fonttitle=\bfseries , colframe=blue!90!black, colback=blue!10!white]
		
		\begin{subfigure}[htbp]{1in}
			\centering
			\scriptsize
			\[
			\Qcircuit @C=0.3em @R=1.5em @!R{
				\lstick{\ket{A}} &	&\ctrl{2}		&\qw	&		&\rstick{\ket{A}}		\\
				\lstick{\ket{B}} &	&\ctrl{1}		&\qw	&  		&\rstick{\ket{B}}		\\
				\lstick{\ket{C}} &	&\targ		&\qw		&		&\rstick{\ket{A \cdot B \oplus C}} 
			}
			\]
			\caption{The Toffoli gate}
			\label{toffoliBAsic}
		\end{subfigure} \qquad 	\begin{subfigure}[htbp]{1.5in}
			\flushleft
			\scriptsize
			\[
			\Qcircuit @C=0.3em @R=2.2em{
				\lstick{\ket{A}} &	&\ctrl{2}		&\qw	&		&\rstick{\ket{A}}	\\
				\lstick{\ket{B}} &	&\qswap \qwx	&\qw	&  	&\rstick{\ket{\overline{A} \cdot B + A \cdot C}}	\\
				\lstick{\ket{C}}&	&\qswap \qwx		&\qw		&		&\rstick{\ket{A \cdot B + \overline{A} \cdot C}}	
			}
			\]
			\caption{The Fredkin gate.}
			\label{fedkinBaisc}
		\end{subfigure}
		\\
		\begin{subfigure}[htbp]{3in}
			
		\end{subfigure}
		\\	   
		\begin{subfigure}[htbp]{3in}
			\flushleft
			\scriptsize
			\[
			\Qcircuit @C=0.3em @R=0.5em @!R{
				\lstick{\ket{A}} &	&\qw &\gate{T}	&\qw		&\targ		&\qw			&\ctrl{2}		&\qw		&\ctrl{1}		&\gate{T^{\dag}}	&\qw			&\ctrl{2}		&\targ		&\qw		& &\rstick{\ket{A}}\\
				\lstick{\ket{B}} &	&\qw		&\gate{T}	&\qw		&\ctrl{-1}		&\targ		&\qw			&\gate{T^{\dag}}	&\targ		&\gate{T^{\dag}}	&\targ		&\qw			&\ctrl{-1}		&\qw		& &\rstick{\ket{B}} \\
				\lstick{\ket{C}} &	&\gate{H}		&\gate{T}	&\qw	&\qw		&\ctrl{-1}		&\targ		&\qw	&\qw		&\gate{T}	&\ctrl{-1}		&\targ		&\gate{H}	&\qw		& &\rstick{\ket{A \cdot B \oplus C}}
			}
			\]
			\caption{Clifford+T implementation of the Toffoli gate.  Source \cite{Maslov}.}
			\label{overview:clifTtoffoli}
		\end{subfigure}
		\\
		\begin{subfigure}[htbp]{2.5in}
			\flushleft
			\scriptsize
			\[
			\Qcircuit @C=0.3em @R=0.7em @!R{
				\lstick{\ket{A}}  &		&\qw &\qw &\gate{T}	&\qw		&\targ		&\qw			&\ctrl{2}		&\qw		&\ctrl{1}		&\gate{T^{\dag}}	&\qw			&\ctrl{2}		&\targ		&\qw		&\qw & &\rstick{\ket{A}}\\
				\lstick{\ket{B}}  &		&\targ &\qw		&\gate{T}	&\qw		&\ctrl{-1}		&\targ		&\qw			&\gate{T^{\dag}}	&\targ		&\gate{T^{\dag}}	&\targ		&\qw			&\ctrl{-1}		&\targ &\qw		& &\rstick{\ket{\overline{A} \cdot B + A \cdot C}}\\
				\lstick{\ket{C}}  &	&\ctrl{-1} &\gate{H}		&\gate{T}	&\qw	&\qw		&\ctrl{-1}		&\targ		&\qw	&\qw		&\gate{T}	&\ctrl{-1}		&\targ		&\gate{H}	&\ctrl{-1} &\qw		& &\rstick{\ket{A \cdot B + \overline{A} \cdot C}}
			}
			\]
			\caption{Clifford+T implementation of the Fredkin gate. Source \cite{Maslov}.}
			\label{overview:clifTfedkin}
		\end{subfigure}
		\caption{The Clifford+T quantum gate implementation of reversible logic gates.}
		\label{Clif-T-reversible}
		
	\end{tcolorbox}
\end{figure}
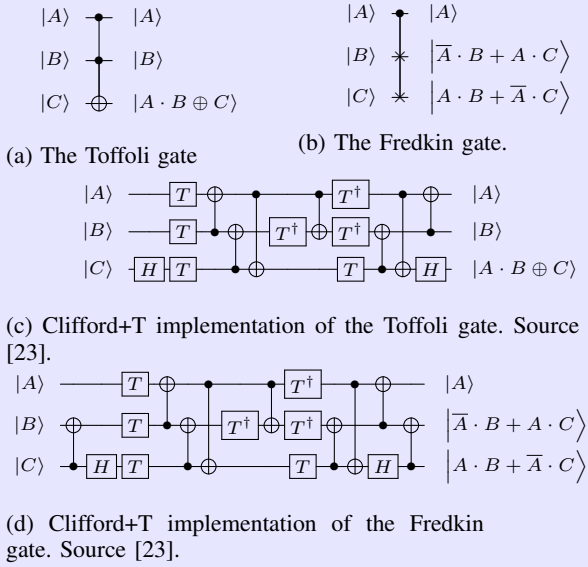

Clifford+T gates can be combined to implement higher level logic gates.  Figure \ref{Clif-T-reversible} illustrates examples of how the Clifford+T gates are used to implement higher level logic gates (specifically the Toffoli gate and the Fredkin gate).  These higher level gates have the functionality shown in Figure \ref{Clif-T-reversible}.  Schematic images also are shown for the Toffoli gate and Fredkin gate.

We focus on the Toffoli and Fredkin gates (see Figure \ref{Clif-T-reversible}) in this article because they are basic building blocks routinely used in the design of reversible logic systems such as those shown in \cite{Edgard2021addition} \cite{Takahashi} \cite{Jayashree} \cite{Edgard2019multiplication}.  These gates are used to construct the quantum circuits presented in this article.

Like the Clifford+T gates, the Toffoli gate and Fredkin gate represent unitary operations and are reversible.  The Toffoli gate takes 3 inputs $A,B,C$ and returns 3 outputs $A,B,A \cdot B \oplus C$.  The Toffoli gate is another example of a controlled gate.  Unlike the CNOT gate, the Toffoli gate has two control qubits.  The result of computation $A \cdot B \oplus C$ is requires that two values ($A$ and $B$ in this case) simultaneously be $1$ before $C$ is complemented.  Thus, the operation performed by the Toffoli gate is identical to the computation performed by the CNOT gate.  Therefore, there are instances were the Toffoli gate is referred to in the literature as a ``doubly controlled CNOT'' gate or as a CCNOT gate.  The Fredkin gate takes 3 inputs $A,B,C$ and returns 3 outputs $A,\overline{A} \cdot B + A \cdot C, A \cdot B + \overline{A} \cdot C $.  So for the Fredkin gate, what is taking place is that when the control input $A = 1$, the values on inputs $B$ and $C$ are interchanged (or swapped).  When $A = 0$, the values at inputs $B$ and $C$ are unchanged.  As a result the Fredkin gate is also referred to as a ``Controlled Swap'' gate or CSWAP gate.

\subsection{Quantum Circuit Resource Cost Measurements}
\label{encycpedia-cost-measure-section}

\begin{figure}[tbhp]
	
	\begin{tcolorbox}[title = Sources of Overhead in Quantum Circuit ,fonttitle=\bfseries , colframe=blue!90!black, colback=blue!10!white]
		\begin{itemize}
			\item \textbf{Garbage Output:}  Garbage outputs are any outputs of a quantum circuit that is not a primary input or a useful output.
			\item \textbf{Ancillae:} Ancillae are additional qubits (often set to a constant value such as $0$) required by a quantum circuit to hold computations.    
			\item \textbf{Qubit Cost:} The total number of qubits required by a quantum circuit. 
			\item \textbf{Depth:} The total number of gate layers in a circuit.  A layer consists of quantum gates operating in parallel.
			\item \textbf{Gate count:} The total number of gates in a quantum circuit. 
		\end{itemize}
		\caption{Cost metrics used in quantum computation.}
		\label{overhead sources-Calout}
	\end{tcolorbox}
	
\end{figure}

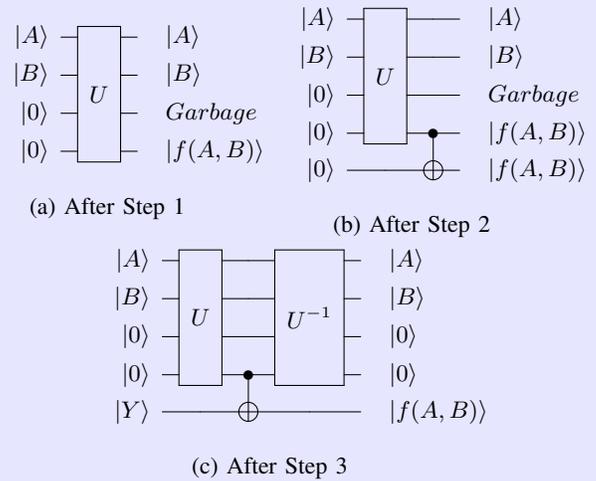
\begin{figure}
	\begin{tcolorbox}[title = Procedure to Remove Garbage Output ,fonttitle=\bfseries , colframe=blue!90!black, colback=blue!10!white]
		\small
		\flushleft
		\begin{subfigure}[thb]{1.3in}
			\flushleft
			\[
			\Qcircuit @C = .7em @R = .7em {
				\lstick{\ket{A}} & \multigate{3}{U} & \qw & \rstick{\ket{A}}\\
				\lstick{\ket{B}} & \ghost{U} & \qw & \rstick{\ket{B}}\\
				\lstick{\ket{0}} & \ghost{U} & \qw & \rstick{Garbage} \\
				\lstick{\ket{0}} & \ghost{U} & \qw & \rstick{\ket{f(A,B)}} \\
			}
			\]
			\caption{After Step 1}
		\end{subfigure}  \qquad \begin{subfigure}[thb]{1.3in}
			
			\[
			\Qcircuit @C = .7em @R = .7em {
				\lstick{\ket{A}} & \multigate{3}{U} & \qw & \qw &\rstick{\ket{A}}\\
				\lstick{\ket{B}} & \ghost{U} & \qw & \qw & \rstick{\ket{B}}\\
				\lstick{\ket{0}} & \ghost{U} & \qw & \qw & \rstick{Garbage} \\
				\lstick{\ket{0}} & \ghost{U} & \ctrl{1}  & \qw & \rstick{\ket{f(A,B)}} \\  
				\lstick{\ket{0}} & \qw &\targ& \qw & \rstick{\ket{f(A,B)}} \\
			}
			\]
			\caption{After Step 2}
		\end{subfigure} \\ \begin{subfigure}[thb]{3in}
			
			\[
			\Qcircuit @C = .7em @R = .7em {
				\lstick{\ket{A}} & \multigate{3}{U} & \qw &\multigate{3}{U^{-1}} & \qw &\rstick{\ket{A}}\\
				\lstick{\ket{B}} & \ghost{U} & \qw & \ghost{U^{-1}} & \qw & \rstick{\ket{B}}\\
				\lstick{\ket{0}} & \ghost{U} & \qw & \ghost{U^{-1}} & \qw & \rstick{\ket{0}} \\
				\lstick{\ket{0}} & \ghost{U} & \ctrl{1} & \ghost{U^{-1}} & \qw & \rstick{\ket{0}} \\  
				\lstick{\ket{Y}} & \qw &\targ& \qw & \qw & \rstick{\ket{f(A,B)}} \\
			}
			\]
			\caption{After Step 3}
		\end{subfigure}
		\caption{General procedure to eliminate garbage outputs produced by a quantum circuit.  $U$ refers to a generic quantum circuit.  To probe further see \cite{Bennett1973trashremoval}.}
		\label{figure:bonusmult}
	\end{tcolorbox}
	
\end{figure}

To determine items such as (i) the suitability of a given quantum circuit design for a given quantum machine or (ii) whether a newly proposed design improves upon the state of the art, we evaluate the resource costs of quantum circuits.  Like in classical computing, we assess a circuit in terms of various resource measures.  Figure \ref{overhead sources-Calout} illustrates sources of resource overhead in a quantum circuit.  Depth (number of gate layers) and total gate count should be familiar as similar measures exist in classical computing (such as critical path delay or transistor count).  Qubit cost refers to the total number of qubits required by a quantum circuit.  Because quantum computation is reversible, ancillae and garbage output are resource overheads which must be considered in quantum circuit design.  Because information cannot be destroyed, there is a need for additional scratch qubits (or ancillae) to hold intermediate steps of computation.  These intermediate steps of computation can end up becoming garbage outputs.  In general, any output that is not part of the original input or a desired output is garbage.  Garbage outputs must be removed via reverse computation before the occupied qubits can be used for subsequent computations.  The general procedure for the removal of garbage outputs is shown in Figure \ref{figure:bonusmult} and outlined in \cite{Bennett1973trashremoval}.

As illustrated in Figure \ref{figure:bonusmult}, a three step procedure is outlined where in \textit{Step 1} the circuit performs its computation, \textit{Step 2} desired outputs are copied to ancillae.  Then, in \textit{Step 3}, the computations in Step 1 are undone.  The result is that qubits holding garbage outputs are now restored to their initial values and can be used for subsequent circuits.  The associated qubits and gates required to clear garbage output add to the overall circuit cost.  The resource cost values reported for a quantum circuit with garbage outputs may not take into account the associated costs of removing garbage outputs.  As a result, resource cost values for a reported for a given circuit may be lower than what is encountered when the circuit is used in computational work.  Thus, the designer should take into account the costs of removing garbage outputs when assessing the suitability of a particular quantum circuit.          

These resource costs measured (gate count, depth, ancillae, garbage output) have been further refined by quantum circuit designers for (i) fault tolerant quantum machines (or FTQ machines)  and (ii) near-term (or noisy intermediate scale quantum NISQ machines).  For the remainder of this Section, we shall examine the resource cost measures for both types of machines

\subsubsection{Resource Cost measures for NISQ machines} 

\begin{figure}[hbtp]
	\centering
	\begin{tcolorbox}[title = Sources of Faults in Quantum Computers ,fonttitle=\bfseries , colframe=blue!90!black, colback=blue!10!white]
		
		\begin{itemize}
			\item \textbf{Gate Error:} Implementing each quantum gate has a probability of incorrect operation.  
			\item \textbf{Energy Relaxation:} A qubit in state $\ket{1}$ decays to state $\ket{0}$ over time.   \item \textbf{Dephasing:} A qubit in superposition loses the superposition state information over time. 
			\item \textbf{State Preparation and Measurement (SPAM) Error:} Qubit initialization and qubit measurement have probabilities of error.  These are reported together since the only way to determine if a qubit was initialized correctly is to measure it.\        
		\end{itemize}
		
		\caption{Sources of error in quantum machines.  Source \cite{Rigetti2019QuantumNISQMachine} \cite{IBM2017QuantumNISQMachine}  \cite{Harty2014QuantumNISQErrorModels} \cite{Alam2019QuantumNISQErrorModels} \cite{Cross2019NISQerror}}
		\label{Prop-table-errors}
		
	\end{tcolorbox}
	
\end{figure}

The existing quantum machines from IBM, Honeywell, IonQ and others are NISQ machines (see \cite{Honeywell2020NISQmachine} \cite{Rigetti2019QuantumNISQMachine} \cite{IBM2017QuantumNISQMachine}).  These quantum machines are large enough to support interesting computations but too small for proposed quantum error correcting codes.  Thus, the choice of resource measures to evaluate NISQ quantum circuits is motivated, in part, by the many sources of error that occur in NISQ machines.  Figure \ref{Prop-table-errors} summarizes some of the reasons why a computation can fail on a NISQ machine.  In addition to non-ideal outcomes from operations such as measurement,
qubits sitting idle can lose their information over time.  This is due to the property of coherence.  While on paper, we can assume fully isolated qubits which can hold state information indefinitely, in practice the information on a qubit shall be disturbed by the environment \cite{Cross2019NISQerror} \cite{Harty2014QuantumNISQErrorModels} \cite{Alam2019QuantumNISQErrorModels}.  Thus, an analogy can be drawn between the qubit and the DRAM cell due to the fact that they both lose their state information over time.

\begin{figure}[tbhp]
	\centering
	\begin{tcolorbox}[title = Resource Cost Measures for NISQ Machines,fonttitle=\bfseries , colframe=blue!90!black, colback=blue!10!white]
		
		\begin{itemize}
			\item Qubit cost: the total number of qubits required to design the quantum circuit. 
			\item CNOT-count: the total number of CNOT gates used in the quantum circuit.  The CNOT gate is shown in Figure \ref{Clifford table}.
			\item CNOT-depth: the maximum number of CNOT gate layers in a quantum circuit.
			\item Garbage output: the number of outputs that exist solely to preserve circuit reversibility. 
			\item $KQ$: the product of the depth and qubit cost.
			\item $KQ_{CNOT}$: The product of the CNOT-depth and qubit cost.   
		\end{itemize}
		
		\caption{ Resource cost metrics appear in works such as: \cite{Kento2020NISQmetrics} \cite{Paler2019NISQerror}}
		\label{NISSQ-cost-measures}
		
	\end{tcolorbox}
	
\end{figure}

Figure \ref{NISSQ-cost-measures} illustrates the measures used to evaluate the resource costs of a quantum circuit implemented on a NISQ machine.  The CNOT gate (see Figure \ref{Clifford table}) has been emphasized because gates that act on two qubits have a higher error rate than one qubit gates \cite{IBM2017QuantumNISQMachine} \cite{Rigetti2019QuantumNISQMachine}  \cite{IonQ2020QuantumNISQMachine}.  A NISQ circuit with many CNOT gates shall face a higher risk of computation failure than a circuit with a reduced CNOT gate count.  Therefore CNOT-count and CNOT-depth are cost measures that are used to evaluate quantum circuit performance \cite{oonishi2020addersqcla}.  The $KQ$ measure is included as it is used in the estimation of the fidelity of a computation on a given quantum circuit \cite{Paler2019NISQerror} \cite{oonishi2020addersqcla}.  The probability of success of a given quantum circuit can be estimated by the following computation:

\begin{tcolorbox}[fonttitle=\bfseries , colframe=green!90!black, colback=green!10!white]
	\begin{equation}
	\frac{1}{K \cdot Q} = A
	\label{eee-encyclopedia-gate-eq-1}
	\end{equation}    
\end{tcolorbox}

\noindent
where $Q$ is the qubit cost and $K$ is the circuit depth.  

A first order estimation of the likelihood of a correct computation of a circuit running on a quantum machine can be achieved by comparing $A$ to the worst case (or largest) failure rate ($\epsilon$) reported for the quantum machine \cite{Paler2019NISQerror}.  If $A < \epsilon$ then a given circuit can deliver correct computations on the NISQ machine.  Conversely, if $A \geq \epsilon$  then the proposed circuit shall require quantum error correcting codes.  $KQ_{CNOT}$ is like the $KQ$ measure in that it is used to estimate fidelity.  However, we only consider the depth of CNOT gates when computing $KQ_{CNOT}$.      

\subsubsection{Resource Cost measures for FTQ machines} 

\begin{figure}[tbhp]
	\centering
	\begin{tcolorbox}[title = Resource Cost Measures for NISQ Machines,fonttitle=\bfseries , colframe=blue!90!black, colback=blue!10!white]
		\begin{itemize}
			\item Qubit cost: the total number of qubits required to design the quantum circuit. 
			\item T-count: the total number of T gates used in the quantum circuit.  The T gate is shown in Figure \ref{Clifford table}.
			\item T-depth: the maximum number of T gate layers in a quantum circuit.
			\item Garbage output: the number of outputs that exist solely to preserve circuit reversibility. 
			\item $KQ_T$: the product of the T-depth and qubit cost.  
		\end{itemize}
		
		\caption{ Resource cost metrics appear in works such as: \cite{Kento2020NISQmetrics} \cite{Mosca2014FaultTolerantQuantumComputing} \cite{Edgard2019divider}. }
		\label{FTQ-cost-measures}
		
	\end{tcolorbox}
\end{figure}

In fault tolerant quantum computation, quantum error correcting codes and fault tolerant gate implementations are used to mitigate the risk of noise errors. While a physical quantum machine that can handle quantum error correcting codes does not exist, researchers have worked to develop fault tolerant gates and circuits  \cite{Miller} \cite{Maslov} \cite{Kiteav2005faulttolerant} \cite{Edgard2019divider} \cite{HaiSheng2020FaultToleantImageCircuits}.  The Clifford+T gate set (see Figure \ref{Clifford table}) has been frequently used in fault tolerant quantum circuit design because it can be made fault tolerant with existing quantum error correcting codes \cite{PalerIOP} \cite{Devitt} \cite{Kiteav2005faulttolerant} \cite{Fowler2008FaultTolerantQuantumComputing} \cite{Bombin} \cite{Fowler2012QuantumSurfaceCodes} \cite{Gosset}.  There are trade offs to using this gate family namely (i) high resource overhead of the fault tolerant implementation of the T gate and (ii) it is approximately universal \cite{Fowler2012QuantumSurfaceCodes} \cite{Haah2017magicstate} \cite{Kliuchnikov2016approximation}.  

In a fault tolerant T gate, ancillae set to the following superposition states $\ket{A}$ and $\ket{Y}$ (where $\ket{Y} = \frac{1}{\sqrt{2}}(\ket{0} + e^{\frac{i \cdot \pi}{2}}\ket{1}$) and  $\ket{A}$ (where $\ket{A} = \frac{1}{\sqrt{2}}(\ket{0} + e^{\frac{i \cdot \pi}{4}}\ket{1})$) are created.  To ensure ancillae are set to $\ket{A}$ and $\ket{Y}$ with sufficient fidelity, a process called state distillation is used.  The state distillation process adds to the overall overhead of the fault tolerant T gate in terms of gate cost, qubit cost and depth \cite{Haah2017magicstate} \cite{Fowler2012QuantumSurfaceCodes}.  The overhead from state distillation results in the T gate overwhelming the other Clifford gates in terms of resource costs \cite{Mosca2014FaultTolerantQuantumComputing}.  As a result, the number of T gates (T-count) and number of T gate layers (T-depth) are used to assess quantum circuit performance.  $KQ_T$ is like the $KQ$ measure in that it is used to estimate fidelity.  However, we only consider the depth of T gates when computing $KQ_{T}$.  

The ``approximately universal'' nature of the Clifford+T gate means that certain types of gates and circuits shall incur prohibitively high T-count and T-depth costs \cite{Kliuchnikov2016approximation}.  This plagues circuits based on gates which perform the following operation:

\begin{tcolorbox}[fonttitle=\bfseries , colframe=green!90!black, colback=green!10!white]
	\begin{equation}
	\begin{bmatrix}
	1 & 0 \\
	0 & e^{\frac{ i \cdot \pi}{n}} 
	\end{bmatrix} 
	\text{ where } n = 2^m
	\label{eee-encyclopedia-gate-eq-2}
	\end{equation}                
\end{tcolorbox}

\noindent
where $m \in \mathbb{Z}$ and $m \geq 3$.   

The high cost of this class of gates is because they cannot be exactly implemented by the Clifford+T gates \cite{Kliuchnikov2013approximationofCliffT} \cite{Kliuchnikov2016approximation}.  To improve the approximation accuracy, more gates are required \cite{Kliuchnikov2013approximationofCliffT}.  This is in stark contrast to the Toffoli and Fredkin gates shown in Figure \ref{Clif-T-reversible} because their associated Clifford+T gate implementations are functionally correct implementations.    Circuit implementations for the quantum Fourier transform and inverse quantum Fourier transform are two examples of quantum circuits based on gates that perform operations shown in equation \ref{eee-encyclopedia-gate-eq-2} where $m \geq 3$.  Due to the importance of these circuits in quantum algorithms (see \cite{Shor1997GeneralNonImageQuantumAlgorithm} \cite{Loke2017QFT} \cite{Fowler2004rotationgates} \cite{HuWenWen2019QuantumWatermarking} \cite{Hales2000QFTalgorithm}), the search for implementations with reduced T-depth and T-count is an active area of research.  Thus, the impact of the approximate universality of the Clifford+T gate set is an important consideration the quantum circuit designer should keep in mind.        

\section{Quantum Circuits for Basic Operations}
\label{Quantum-circuit-arithmetic}

Quantum circuits for elementary functions such as arithmetic functions are used as building blocks in implementations of quantum algorithms.  In this Section we present examples of quantum circuits used for basic arithmetic which are used as building blocks in larger quantum datapath circuits such as those presented in \cite{VanDam2008exponential} \cite{Edgard2018JETCsqrt} \cite{Shor1997GeneralNonImageQuantumAlgorithm} \cite{Edgard2018bilinear} \cite{Quipper} \cite{Soeken2017reciprocal} \cite{Bhaskar2016QuantumScientificComputation}.  As a result, the design of quantum circuits for arithmetic functions is an active area of research and quantum programming languages such as Quipper \cite{Quipper} and $LIQUi\vert\rangle$ \cite{LIQUi} include libraries of quantum arithmetic circuits.       

This Section shall present circuits for several basic arithmetic functions.  The circuits shown in Sections \ref{addition-ee-encyclopedia} \ref{subtraction-ee-encyclopedua} \ref{multiplication-section} and \ref{ee-wiley-division} are examples of quantum circuits which accept input that are either (i) a single boolean value or (ii) multiple boolean values in superposition.  These circuits perform computations analogous to classical arithmetic circuits and return results of computation that are (i) a single boolean value or (ii) multiple boolean values in superposition.  As we have learned, quantum gates can return superposition states.  Thus, it is possible to represent computations in terms of superposition state manipulations.  For these circuits, at the end of computation, the result shall appear as a modification to the original superposition state of the input.  An example is the quantum Fourier transform which was presented earlier (see equation \ref{encyclopedia-ee-eq7}).  To acquaint you with this class of circuit, Section \ref{QFT} illustrates an example of the quantum circuit implementation of the quantum Fourier transform.

\subsection{Addition}
\label{addition-ee-encyclopedia}

\begin{figure}[tbhp]
	\centering
	\begin{tcolorbox}[title = Quantum Addition Circuit and its Building Blocks,fonttitle=\bfseries , colframe=blue!90!black, colback=blue!10!white]
		
		\begin{subfigure}[hbtp]{1.25in}
			\scriptsize
			\[
			\Qcircuit @C = .4em @R = .7em @!R{
				\lstick{\ket{A}} & \targ 		& \ctrl{2} & \qw &   & & & & & & & & & & & & &  \lstick{\ket{A}} & \multigate{2}{\begin{sideways} MAJ \end{sideways}} & \qw \\ 
				\lstick{\ket{B}} & \targ		& \ctrl{1} & \qw &  & & = & & & & & & & & & & & \lstick{\ket{B}} & \ghost{\begin{sideways} MAJ \end{sideways}} & \qw \\
				\lstick{\ket{C}} & \ctrl{-2}	& \targ    & \qw &  & & & & & & & & & & & & &   \lstick{\ket{C}} & \ghost{\begin{sideways} MAJ \end{sideways}} & \qw
			}
			\]
			\caption{ Majority (MAJ) building block and its circuit implementation.  Taken from \cite{Cuccaro2004adder} }
		\end{subfigure}	\qquad \begin{subfigure}[hbtp]{1.25in}	
			\scriptsize
			\[
			\Qcircuit @C = .4em @R = .7em @!R{
				\lstick{\ket{A}} & \ctrl{2} & \targ 	& \ctrl{1} & \qw &  & & & & & & & & & & & & &  \lstick{\ket{A}} & \multigate{2}{\begin{sideways} UMA \end{sideways}} & \qw  \\
				\lstick{\ket{B}} & \ctrl{1} & \qw 		& \targ	   & \qw & & & = & & & & & & & & & & &  \lstick{\ket{B}} & \ghost{\begin{sideways} UMA \end{sideways}} & \qw\\
				\lstick{\ket{C}} & \targ    & \ctrl{-2} & \qw 	   & \qw & & & & & & & & & & & & & &  \lstick{\ket{C}} & \ghost{\begin{sideways} UMA \end{sideways}} & \qw	
			}
			\]
			\caption{Unmajority and Add (UMA) building block and its circuit implementation.  Taken from \cite{Cuccaro2004adder} }
		\end{subfigure}
		\\
		\begin{subfigure}[hbtp]{3in}
			\[
			\Qcircuit @C = .5em @R = .7em @!R{
				\lstick{\ket{0}}   & \multigate{2}{\begin{sideways} MAJ \end{sideways}} & \qw & \qw &   \qw &  \qw &   \qw & \rstick{\ket{C_0 \oplus A_0}} \\
				\lstick{\ket{B_0}} & \ghost{\begin{sideways} MAJ \end{sideways}} 		 & \qw & \qw &  \qw &  \qw &   \qw & \rstick{\ket{B_0 \oplus A_0}} \\
				\lstick{\ket{A_0}} & \ghost{\begin{sideways} MAJ \end{sideways}} 		 &  \multigate{2}{\begin{sideways} MAJ \end{sideways}} & \qw &  \qw &  \qw &   \qw & \rstick{\ket{C_1 \oplus A_1}} \\
				\lstick{\ket{B_1}} & \qw & \ghost{\begin{sideways} MAJ \end{sideways}} 		 & \qw & \qw &  \qw &   \qw & \rstick{\ket{B_1 \oplus A_1}} \\
				\lstick{\ket{A_1}} & \qw & \ghost{\begin{sideways} MAJ \end{sideways}} 		 & \multigate{2}{\begin{sideways} MAJ \end{sideways}} & \qw &  \qw &   \qw & \rstick{\ket{C_2 \oplus A_2}} \\
				\lstick{\ket{B_2}} & \qw & \qw & \ghost{\begin{sideways} MAJ \end{sideways}} 		 & \qw &  \qw &   \qw & \rstick{\ket{B_2 \oplus A_2}} \\
				\lstick{\ket{A_2}} & \qw & \qw & \ghost{\begin{sideways} MAJ \end{sideways}} 		 & \multigate{2}{\begin{sideways} MAJ \end{sideways}} &  \qw &   \qw & \rstick{\ket{C_3 \oplus A_3}} \\
				\lstick{\ket{B_3}} & \qw & \qw & \qw & \ghost{\begin{sideways} MAJ \end{sideways}} &  \qw &   \qw & \rstick{\ket{B_3 \oplus A_3}} \\
				\lstick{\ket{A_3}} & \qw & \qw & \qw & \ghost{\begin{sideways} MAJ \end{sideways}} & \ctrl{1} &\qw & \rstick{\ket{C_4 }} \\
				\lstick{\ket{0}}   & \qw & \qw & \qw & \qw & \targ & \qw & \rstick{\ket{C_4 }} \\	 
			}
			\]
			\caption{Implementation of a quantum addition circuit.  Synthesis of carry bits.  Taken from \cite{Cuccaro2004adder}. }
			\label{adder-halfway}
		\end{subfigure}
		\\
		\begin{subfigure}[hbtp]{3in}
			\centering
			\[
			\Qcircuit @C = .3em @R = .7em @!R{
				\lstick{\ket{0}}   & \multigate{2}{\begin{sideways} MAJ \end{sideways}} & \qw &   \qw &  \qw &   \qw &  \qw &  \qw &  \qw & \multigate{2}{\begin{sideways} UMA \end{sideways}} &\qw & \rstick{\ket{0}} \\
				\lstick{\ket{B_0}} & \ghost{\begin{sideways} MAJ \end{sideways}} 		 & \qw &  \qw &  \qw &   \qw &  \qw &  \qw &  \qw & \ghost{\begin{sideways} UMA \end{sideways}} &\qw & \rstick{\ket{S_0}} \\
				\lstick{\ket{A_0}} & \ghost{\begin{sideways} MAJ \end{sideways}} 		 &  \multigate{2}{\begin{sideways} MAJ \end{sideways}} & \qw &  \qw &   \qw &  \qw &  \qw & \multigate{2}{\begin{sideways} UMA \end{sideways}} &  \ghost{\begin{sideways} UMA \end{sideways}} &\qw & \rstick{\ket{A_0}} \\
				\lstick{\ket{B_1}} & \qw & \ghost{\begin{sideways} MAJ \end{sideways}} 		 & \qw &  \qw &   \qw &  \qw &  \qw & \ghost{\begin{sideways} UMA \end{sideways}} &  \qw & \qw & \rstick{\ket{S_1 }} \\
				\lstick{\ket{A_1}} & \qw & \ghost{\begin{sideways} MAJ \end{sideways}} 		 & \multigate{2}{\begin{sideways} MAJ \end{sideways}} & \qw &  \qw &  \qw & \multigate{2}{\begin{sideways} UMA \end{sideways}} & \ghost{\begin{sideways} UMA \end{sideways}} &  \qw & \qw & \rstick{\ket{A_1}} \\
				\lstick{\ket{B_2}} & \qw & \qw & \ghost{\begin{sideways} MAJ \end{sideways}} 		 & \qw &  \qw &\qw & \ghost{\begin{sideways} UMA \end{sideways}} & \qw &\qw &\qw & \rstick{\ket{S_2}} \\
				\lstick{\ket{A_2}} & \qw & \qw & \ghost{\begin{sideways} MAJ \end{sideways}} 		 & \multigate{2}{\begin{sideways} MAJ \end{sideways}} &  \qw & \multigate{2}{\begin{sideways} UMA \end{sideways}} & \ghost{\begin{sideways} UMA \end{sideways}} &  \qw & \qw &\qw & \rstick{\ket{A_2}} \\
				\lstick{\ket{B_3}} & \qw & \qw & \qw & \ghost{\begin{sideways} MAJ \end{sideways}} &  \qw & \ghost{\begin{sideways} UMA \end{sideways}}&  \qw &  \qw & \qw & \qw & \rstick{\ket{S_3}} \\
				\lstick{\ket{A_3}} & \qw & \qw & \qw & \ghost{\begin{sideways} MAJ \end{sideways}} & \ctrl{1} & \ghost{\begin{sideways} UMA \end{sideways}} &  \qw &  \qw & \qw &\qw & \rstick{\ket{A_3 }} \\
				\lstick{\ket{0}}   & \qw & \qw & \qw & \qw & \targ & \qw &  \qw &  \qw & \qw & \qw & \rstick{\ket{S_4 }} \\	 
			}
			\]
			\caption{Implementation of a quantum addition circuit. Taken from \cite{Cuccaro2004adder}}
			\label{adder-complete}
		\end{subfigure}
		
		\caption{Quantum circuit for addition. }
		\label{adder-unit}
		
	\end{tcolorbox}
	
\end{figure}

Figure \ref{adder-unit} illustrates an example of a quantum circuit for addition that was first presented in \cite{Cuccaro2004adder}.  As opposed to alternative architectures such as those shown in \cite{RiGuoZhou2017ImageInterpolation}, this style of addition circuit enjoys no garbage outputs.  

Consider the conditional addition of two $n$-bit numbers $a$ and $b$ which are stored in quantum registers $\ket{A} $ and $\ket{B}$.  A quantum register is an $n$ bit array of qubits.  Two ancillae are used (see Figure \ref{adder-unit}).  One if set to $\ket{0}$ and shall have the most significant bit of the sum $s_n$.  The second ancillae can be set to $\ket{0}$ or hold a carry in bit $c_0$. At the end of computation, quantum register $\ket{B}$ shall contain the sum bits $s_0$ through $s_{n-1}$.  Quantum register $\ket{A}$ shall have the value $a$ and the ancillae originally set to $\ket{0}$ or $c_0$ shall be restored to its original value.

This circuit cleverly exploits the Bennett's garbage removal method presented in \cite{Bennett1973trashremoval} and shown in Figure \ref{figure:bonusmult} to eliminate garbage output.  Each carry bit is generated first.  This is accomplished by the array of Majority (\textit{MAJ}) circuit blocks (the construction of a \textit{MAJ} block in terms of reversible logic gates is shown in Figure \ref{adder-unit}).  The most significant carry bit $c_n$ is then copied to an ancillae because $c_n = s_n$.  Hence the CNOT gate which takes the qubit with the value $c_n$ and an ancillae as inputs appears in Figure \ref{adder-unit}.  Then the computation of the carries is reversed in a manner such that the remaining sum bits are produced and any intermediate computations are removed.  This uncomputation is the work performed by the array of Unmajority and Add (\textit{UMA}) circuit blocks (the construction of a \textit{UMA} block in terms of reversible logic gates is shown in Figure \ref{adder-unit}).  Note that the \textit{UMA} blocks are applied in reverse order of the \textit{MAJ} blocks.  This is intentional and is because we need to uncompute the carry bits by first starting with the most significant bit $c_n$ and working backward toward the least significant carry bit $c_0$.      

The design of addition circuit continues and alternative designs free of garbage outputs have been proposed such as \cite{oonishi2020addersqcla} \cite{Lin} \cite{Saeedi2} \cite{Thapliyal2013adder} \cite{Edgard2021addition} that offer reductions in cost measures such as T-count or qubit costs.  Useful variants such as conditional adders (adders whose functionality can be disabled via additional control signals) are being developed as well \cite{Lin} \cite{Jayashree} \cite{Edgard2019multiplication}.              

\subsection{Subtraction}
\label{subtraction-ee-encyclopedua}

\begin{figure}[tbhp]
	\centering
	\begin{tcolorbox}[title = Quantum Subtraction Circuit,fonttitle=\bfseries , colframe=blue!90!black, colback=blue!10!white]
		\[
		\Qcircuit @C = .3em @R = .7em @!R {
			\lstick{\ket{0}}   &\qw &  \multigate{2}{\begin{sideways} MAJ \end{sideways}} & \qw &   \qw &  \qw &   \qw &  \qw &  \qw &  \qw & \multigate{2}{\begin{sideways} UMA \end{sideways}} &\qw &\qw & \rstick{\ket{0}} \\
			\lstick{\ket{B_0}} &\targ & \ghost{\begin{sideways} MAJ \end{sideways}} 		 & \qw &  \qw &  \qw &   \qw &  \qw &  \qw &  \qw & \ghost{\begin{sideways} UMA \end{sideways}} &\targ &\qw & \rstick{\ket{S_0}} \\
			\lstick{\ket{A_0}} &\qw &  \ghost{\begin{sideways} MAJ \end{sideways}} 		 &  \multigate{2}{\begin{sideways} MAJ \end{sideways}} & \qw &  \qw &   \qw &  \qw &  \qw & \multigate{2}{\begin{sideways} UMA \end{sideways}} &  \ghost{\begin{sideways} UMA \end{sideways}} &\qw &\qw & \rstick{\ket{A_0}} \\
			\lstick{\ket{B_1}} &\targ & \qw & \ghost{\begin{sideways} MAJ \end{sideways}} 		 & \qw &  \qw &   \qw &  \qw &  \qw & \ghost{\begin{sideways} UMA \end{sideways}} &  \qw &\targ & \qw & \rstick{\ket{S_1 }} \\
			\lstick{\ket{A_1}} &\qw & \qw & \ghost{\begin{sideways} MAJ \end{sideways}} 		 & \multigate{2}{\begin{sideways} MAJ \end{sideways}} & \qw &  \qw &  \qw & \multigate{2}{\begin{sideways} UMA \end{sideways}} & \ghost{\begin{sideways} UMA \end{sideways}} &  \qw &\qw & \qw & \rstick{\ket{A_1}} \\
			\lstick{\ket{B_2}} &\targ & \qw & \qw & \ghost{\begin{sideways} MAJ \end{sideways}} 		 & \qw &  \qw &\qw & \ghost{\begin{sideways} UMA \end{sideways}} & \qw &\qw &\targ &\qw & \rstick{\ket{S_2}} \\
			\lstick{\ket{A_2}} &\qw & \qw & \qw & \ghost{\begin{sideways} MAJ \end{sideways}} 		 & \multigate{2}{\begin{sideways} MAJ \end{sideways}} &  \qw & \multigate{2}{\begin{sideways} UMA \end{sideways}} & \ghost{\begin{sideways} UMA \end{sideways}} &  \qw & \qw &\qw &\qw & \rstick{\ket{A_2}} \\
			\lstick{\ket{B_3}} &\targ & \qw & \qw & \qw & \ghost{\begin{sideways} MAJ \end{sideways}} &  \qw & \ghost{\begin{sideways} UMA \end{sideways}}&  \qw &  \qw & \qw &\targ & \qw & \rstick{\ket{S_3}} \\
			\lstick{\ket{A_3}} &\qw & \qw & \qw & \qw & \ghost{\begin{sideways} MAJ \end{sideways}} & \ctrl{1} & \ghost{\begin{sideways} UMA \end{sideways}} &  \qw &  \qw & \qw &\qw &\qw & \rstick{\ket{A_3 }} \\
			\lstick{\ket{0}}   &\qw & \qw & \qw & \qw & \qw & \targ & \qw &  \qw &  \qw & \qw & \qw &\qw & \rstick{\ket{S_4 }} \\	 
		}
		\]
		
		\caption{Implementation of a quantum subtraction circuit based on the addition circuit presented in \cite{Cuccaro2004adder}.}
		\label{subtraction-complete}
	\end{tcolorbox}
\end{figure}

\begin{figure}[tbhp]
	\centering
	\begin{tcolorbox}[title = Guide to Convert an Addition Circuit to Perform Other Arithmetic Functions,fonttitle=\bfseries , colframe=blue!90!black, colback=blue!10!white]
		
		\begin{subfigure}[hbtp]{1.25in}
			\centering	
			\[
			\Qcircuit @C = .4em @R = .7em @!R {
				\lstick{\ket{B_0}} &\targ & \multigate{7}{\begin{sideways} ADDITION \end{sideways}} & \targ & \qw & &\rstick{\ket{D_0}} \\
				\lstick{\ket{A_0}} &\qw & \ghost{\begin{sideways} ADDITION \end{sideways}} & \qw & \qw & &\rstick{\ket{A_0}} \\
				\lstick{\ket{B_1}} &\targ & \ghost{\begin{sideways} ADDITION \end{sideways}} &  \targ & \qw & &\rstick{\ket{D_1}} \\
				\lstick{\ket{A_1}} &\qw & \ghost{\begin{sideways} ADDITION \end{sideways}} & \qw & \qw & &\rstick{\ket{A_1}} \\
				\lstick{\ket{B_2}} &\targ & \ghost{\begin{sideways} ADDITION \end{sideways}} &  \targ & \qw & &\rstick{\ket{D_2}} \\
				\lstick{\ket{A_2}} &\qw & \ghost{\begin{sideways} ADDITION \end{sideways}} & \qw & \qw & &\rstick{\ket{A_2}} \\
				\lstick{\ket{B_3}} &\targ & \ghost{\begin{sideways} ADDITION \end{sideways}} &  \targ & \qw & &\rstick{\ket{D_3}} \\
				\lstick{\ket{A_3}} &\qw & \ghost{\begin{sideways} ADDITION \end{sideways}} & \qw & \qw & &\rstick{\ket{A_3}} \\
			}
			\]	
			\caption{How to modify an addition circuit to perform subtraction.  Source: \cite{Thapliyal2016addsub}}
		\end{subfigure} \qquad \begin{subfigure}[hbtp]{1.35in}
			\centering
			\[
			\Qcircuit @C = .4em @R = .7em @!R {
				\lstick{\ket{Ctrl}} &\ctrl{7} & \qw &\ctrl{7} & \qw &\rstick{\ket{Ctrl}} \\ 
				\lstick{\ket{B_0}} &\targ & \multigate{7}{\begin{sideways} ADDITION \end{sideways}} & \targ & \qw & &\rstick{\ket{S_0}} \\
				\lstick{\ket{A_0}} &\qw & \ghost{\begin{sideways} ADDITION \end{sideways}} & \qw & \qw & &\rstick{\ket{A_0}} \\
				\lstick{\ket{B_1}} &\targ & \ghost{\begin{sideways} ADDITION \end{sideways}} &  \targ & \qw & &\rstick{\ket{S_1}} \\
				\lstick{\ket{A_1}} &\qw & \ghost{\begin{sideways} ADDITION \end{sideways}} & \qw & \qw & &\rstick{\ket{A_1}} \\
				\lstick{\ket{B_2}} &\targ & \ghost{\begin{sideways} ADDITION \end{sideways}} &  \targ & \qw & &\rstick{\ket{S_2}} \\
				\lstick{\ket{A_2}} &\qw & \ghost{\begin{sideways} ADDITION \end{sideways}} & \qw & \qw & &\rstick{\ket{A_2}} \\
				\lstick{\ket{B_3}} &\targ & \ghost{\begin{sideways} ADDITION \end{sideways}} &  \targ & \qw & &\rstick{\ket{S_3}} \\
				\lstick{\ket{A_3}} &\qw & \ghost{\begin{sideways} ADDITION \end{sideways}} & \qw & \qw & &\rstick{\ket{A_3}} \\
			}
			\]	
			\caption{How to modify an addition circuit so that it can perform addition or subtraction.  Source: \cite{Thapliyal2016addsub}}
		\end{subfigure}
		
		\caption{ Circuits for subtraction and conditional addition or subtraction  }
		\label{subtraction-unit-ee-encyclopedia}
	\end{tcolorbox}
\end{figure}

In classical computation, subtraction can be accomplished with arrays of full subtraction blocks or by modifying an addition circuit so that it performs subtraction.  In this article, we shall focus on the latter approach.  Figure \ref{subtraction-complete} shows an implementation of an adder proposed in \cite{Thapliyal2016addsub} that has been modified to perform subtraction.  The adder used is the same design as the one shown in Figure \ref{adder-unit}.  Figure \ref{subtraction-unit-ee-encyclopedia} shows how a generic addition circuit can be made to perform subtraction.

Consider the subtraction of two $n$-bit numbers $a$ and $b$ which are stored in quantum registers $\ket{A} $ and $\ket{B}$.  The subtraction circuit calculates $(\overline{\bar{\ket{B}}+\ket{A})}$ which simplifies to $\ket{B}-\ket{A}$  \cite{Thapliyal2016addsub} \cite{Edgard2019divider}. At the end of computation, $\ket{A}$ has the input value $a$ and $\ket{B}$ contains the difference $d$.  As shown in Figure \ref{subtraction-unit-ee-encyclopedia}, by adding arrays of NOT gates to the before and after the quantum addition circuit, the whole circuit performs subtraction.  The functionality of the quantum adder used is not impacted by the additional hardware.     

The cost of this particular circuit is a function of the underlying addition circuit used.  Thus, as researchers present more resource efficient addition circuits, the cost of subtraction circuits shall also stand to benefit.

\subsection{Multiplication}
\label{multiplication-section}

\begin{figure}[tbhp]
	\centering
	\begin{tcolorbox}[title = Quantum Multiplication Circuit,fonttitle=\bfseries , colframe=blue!90!black, colback=blue!10!white]
		\[
		\Qcircuit @C = .7em @R = .7em @!R {
			\lstick{\ket{B_4}} & \ctrl{5} & \qw &   \qw &   \qw &   \qw &   \qw & &\rstick{\ket{B_4}}\\
			\lstick{\ket{B_3}} & \qw & \ctrl{5} & \qw &   \qw &   \qw &   \qw &  &\rstick{\ket{B_3}}\\
			\lstick{\ket{B_2}} & \qw & \qw & \ctrl{5} & \qw &   \qw &   \qw & &\rstick{\ket{B_2}} \\
			\lstick{\ket{B_1}}  & \qw & \qw & \qw & \ctrl{5} & \qw &   \qw &  &\rstick{\ket{B_1}} \\
			\lstick{\ket{B_0}} &\qw & \qw & \qw & \qw & \ctrl{5} & \qw & &\rstick{\ket{B_0}}  \\ 
			\lstick{\ket{A_{[4:0]}}} & \ctrl{1} &  \ctrl{2} &  \ctrl{3} &  \ctrl{4} &  \ctrl{5} &  \qw & \rstick{\ket{A_{[4:0]}}}\\
			\lstick{\ket{0}} & \multigate{4}{\begin{sideways} TGA \end{sideways}} & \qw & \qw & \qw & \qw & \qw & &\rstick{\ket{P_0}}\\
			\lstick{\ket{0}} & \ghost{\begin{sideways} TGA \end{sideways}} & \multigate{6}{\begin{sideways} Conditional Add \end{sideways}} & \qw & \qw & \qw & \qw & &\rstick{\ket{P_1}}\\
			\lstick{\ket{0}} & \ghost{\begin{sideways} TGA \end{sideways}} & \ghost{\begin{sideways} Conditional Add \end{sideways}} & \multigate{6}{\begin{sideways} Conditional Add \end{sideways}} &  \qw & \qw & \qw & &\rstick{\ket{P_2}} \\
			\lstick{\ket{0}} & \ghost{\begin{sideways} TGA \end{sideways}} & \ghost{\begin{sideways} Conditional Add \end{sideways}} & \ghost{\begin{sideways} Conditional Add \end{sideways}} & \multigate{6}{\begin{sideways} Conditional Add \end{sideways}} & \qw & \qw & &\rstick{\ket{P_3}} \\ 
			\lstick{\ket{0}} & \ghost{\begin{sideways} TGA \end{sideways}} & \ghost{\begin{sideways} Conditional Add \end{sideways}} & \ghost{\begin{sideways} Conditional Add \end{sideways}} & \ghost{\begin{sideways} Conditional Add \end{sideways}} & \multigate{6}{\begin{sideways} Conditional Add \end{sideways}} &  \qw & &\rstick{\ket{P_4}}\\
			\lstick{\ket{0}} & \qw & \ghost{\begin{sideways} Conditional Add \end{sideways}} & \ghost{\begin{sideways} Conditional Add \end{sideways}} & \ghost{\begin{sideways} Conditional Add \end{sideways}} & \ghost{\begin{sideways} Conditional Add \end{sideways}} & \qw & &\rstick{\ket{P_5}}\\
			\lstick{\ket{0}} & \qw & \ghost{\begin{sideways} Conditional Add \end{sideways}} & \ghost{\begin{sideways} Conditional Add \end{sideways}} & \ghost{\begin{sideways} Conditional Add \end{sideways}} & \ghost{\begin{sideways} Conditional Add \end{sideways}} & \qw & \ &\rstick{\ket{P_6}}\\
			\lstick{\ket{0}} & \qw & \ghost{\begin{sideways} Conditional Add \end{sideways}} & \ghost{\begin{sideways} Conditional Add \end{sideways}} & \ghost{\begin{sideways} Conditional Add \end{sideways}} & \ghost{\begin{sideways} Conditional Add \end{sideways}} & \qw &  &\rstick{\ket{P_7}}\\
			\lstick{\ket{0}} & \qw &  \qw & \ghost{\begin{sideways} Conditional Add \end{sideways}} &  \ghost{\begin{sideways} Conditional Add \end{sideways}} & \ghost{\begin{sideways} Conditional Add \end{sideways}} & \qw &  &\rstick{\ket{P_8}}\\
			\lstick{\ket{0}} & \qw & \qw &	\qw & \ghost{\begin{sideways} Conditional Add \end{sideways}} & \ghost{\begin{sideways} Conditional Add \end{sideways}} & \qw &  &\rstick{\ket{P_9}}\\
			\lstick{\ket{0}} & \qw & \qw & \qw & \qw & \ghost{\begin{sideways} Conditional Add \end{sideways}} & \qw &  &\rstick{\ket{0}}\\
		}
		\]
		\caption{Quantum circuit for the multiplication of two $5$ bit values.  The blocks labeled ``Conditional Add'' are conditional addition circuits and block labeled ``TGA'' is an array of Toffoli gates.  Adapted from the design shown in \cite{Edgard2019multiplication}. }
		\label{ee-multiplication-circuit}
	\end{tcolorbox}
\end{figure}
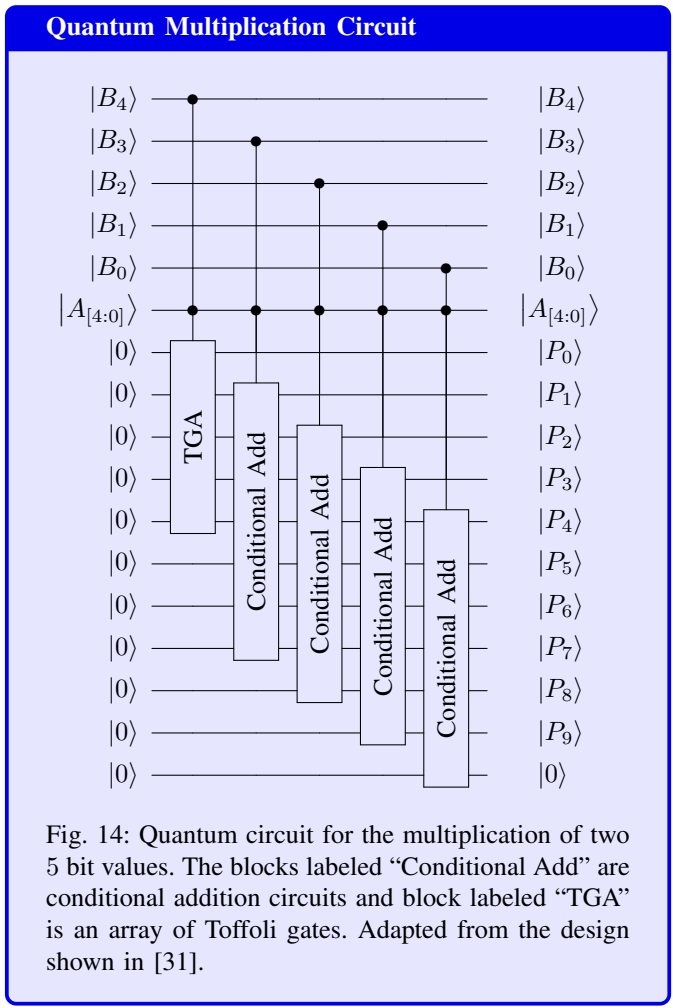

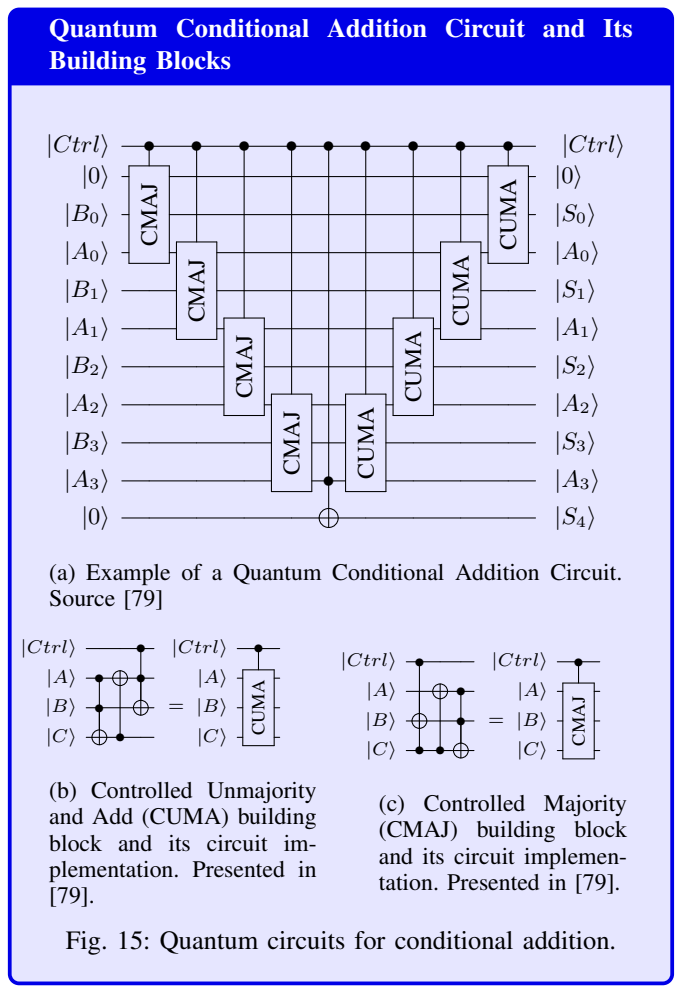
\begin{figure}[tbhp]
	\centering
	\begin{tcolorbox}[title = Quantum Conditional Addition Circuit and Its Building Blocks,fonttitle=\bfseries , colframe=blue!90!black, colback=blue!10!white]

		\begin{subfigure}[hbtp]{3in}
			\centering
			\small
			\[
			\Qcircuit @C = .3em @R = .7em {
				\lstick{\ket{Ctrl}} & \ctrl{1} & \ctrl{3} & \ctrl{5} & \ctrl{7} &\ctrl{9} & \ctrl{7} & \ctrl{5} & \ctrl{3} & \ctrl{1} &\qw & &\rstick{\ket{Ctrl}} \\
				\lstick{\ket{0}}   & \multigate{2}{\begin{sideways} CMAJ \end{sideways}} & \qw &   \qw &  \qw &   \qw &  \qw &  \qw &  \qw & \multigate{2}{\begin{sideways} CUMA \end{sideways}} &\qw & \rstick{\ket{0}} \\
				\lstick{\ket{B_0}} & \ghost{\begin{sideways} CMAJ \end{sideways}} 		 & \qw &  \qw &  \qw &   \qw &  \qw &  \qw &  \qw & \ghost{\begin{sideways} CUMA \end{sideways}} &\qw & \rstick{\ket{S_0}} \\
				\lstick{\ket{A_0}} & \ghost{\begin{sideways} CMAJ \end{sideways}} 		 &  \multigate{2}{\begin{sideways} CMAJ \end{sideways}} & \qw &  \qw &   \qw &  \qw &  \qw & \multigate{2}{\begin{sideways} CUMA \end{sideways}} &  \ghost{\begin{sideways} CUMA \end{sideways}} &\qw & \rstick{\ket{A_0}} \\
				\lstick{\ket{B_1}} & \qw & \ghost{\begin{sideways} CMAJ \end{sideways}} 		 & \qw &  \qw &   \qw &  \qw &  \qw & \ghost{\begin{sideways} CUMA \end{sideways}} &  \qw & \qw & \rstick{\ket{S_1 }} \\
				\lstick{\ket{A_1}} & \qw & \ghost{\begin{sideways} CMAJ \end{sideways}} 		 & \multigate{2}{\begin{sideways} CMAJ \end{sideways}} & \qw &  \qw &  \qw & \multigate{2}{\begin{sideways} CUMA \end{sideways}} & \ghost{\begin{sideways} CUMA \end{sideways}} &  \qw & \qw & \rstick{\ket{A_1}} \\
				\lstick{\ket{B_2}} & \qw & \qw & \ghost{\begin{sideways} CMAJ \end{sideways}} 		 & \qw &  \qw &\qw & \ghost{\begin{sideways} CUMA \end{sideways}} & \qw &\qw &\qw & \rstick{\ket{S_2}} \\
				\lstick{\ket{A_2}} & \qw & \qw & \ghost{\begin{sideways} CMAJ \end{sideways}} 		 & \multigate{2}{\begin{sideways} CMAJ \end{sideways}} &  \qw & \multigate{2}{\begin{sideways} CUMA \end{sideways}} & \ghost{\begin{sideways} CUMA \end{sideways}} &  \qw & \qw &\qw & \rstick{\ket{A_2}} \\
				\lstick{\ket{B_3}} & \qw & \qw & \qw & \ghost{\begin{sideways} CMAJ \end{sideways}} &  \qw & \ghost{\begin{sideways} CUMA \end{sideways}}&  \qw &  \qw & \qw & \qw & \rstick{\ket{S_3}} \\
				\lstick{\ket{A_3}} & \qw & \qw & \qw & \ghost{\begin{sideways} CMAJ \end{sideways}} & \ctrl{1} & \ghost{\begin{sideways} CUMA \end{sideways}} &  \qw &  \qw & \qw &\qw & \rstick{\ket{A_3 }} \\
				\lstick{\ket{0}}   & \qw & \qw & \qw & \qw & \targ & \qw &  \qw &  \qw & \qw & \qw & \rstick{\ket{S_4 }} \\	 
			}
			\]
			\caption{Example of a Quantum Conditional Addition Circuit.  Source \cite{Lin} }
		\end{subfigure}
		\\
		\begin{subfigure}[tbhp]{1.4in}
			\centering
			\scriptsize
			\[
			\Qcircuit @C = .3em @R = .7em @!R{
				\lstick{\ket{Ctrl}} & \qw &\qw &\ctrl{2} &\qw & & & & & & & & & & & & & & &  \lstick{\ket{Ctrl}} & \ctrl{1} & \qw \\ 
				\lstick{\ket{A}} & \ctrl{2} & \targ 	& \ctrl{1} & \qw &  &  & & & & & & & & & & & & &   \lstick{\ket{A}} & \multigate{2}{\begin{sideways} CUMA \end{sideways}} & \qw  \\
				\lstick{\ket{B}} & \ctrl{1} & \qw 		& \targ	   & \qw & & & & = & & & & & & & & & & &  \lstick{\ket{B}} & \ghost{\begin{sideways} CUMA \end{sideways}} & \qw\\
				\lstick{\ket{C}} & \targ    & \ctrl{-2} & \qw 	   & \qw & & & & &  & & & & & & & & & &  \lstick{\ket{C}} & \ghost{\begin{sideways} CUMA \end{sideways}} & \qw	
			}
			\]
			\caption{Controlled Unmajority and Add (CUMA) building block and its circuit implementation.  Presented in \cite{Lin}.}
			\label{CUMA-circuit-ee}
		\end{subfigure} \qquad \begin{subfigure}[tbhp]{1.3in}
			\centering
			\scriptsize
			\[
			\Qcircuit @C = .3em @R = .7em @!R{
				\lstick{\ket{Ctrl}} & \ctrl{2} &\qw &\qw &\qw & & & & & & & & & & & & & & &  \lstick{\ket{Ctrl}} & \ctrl{1} &\qw \\ 
				\lstick{\ket{A}} & \qw &\targ 		& \ctrl{2} & \qw &   & & & & & & & & & & & & & &  \lstick{\ket{A}} & \multigate{2}{\begin{sideways} CMAJ \end{sideways}} & \qw \\ 
				\lstick{\ket{B}} & \targ	&\qw	& \ctrl{1} & \qw &  & & & = & & & & & & & & & & &  \lstick{\ket{B}} & \ghost{\begin{sideways} CMAJ \end{sideways}} & \qw \\
				\lstick{\ket{C}} & \ctrl{-2} & \ctrl{-2}	& \targ    & \qw &  & & & & & & & & & & & & & &  \lstick{\ket{C}} & \ghost{\begin{sideways} CMAJ \end{sideways}} & \qw
			}
			\]
			\caption{Controlled Majority (CMAJ) building block and its circuit implementation.  Presented in \cite{Lin}.}
			\label{CMAJ-circuit-ee}
		\end{subfigure}
		\caption{ Quantum circuits for conditional addition.  }
		\label{Conditional-Add-Quantum-ee}
	\end{tcolorbox}
\end{figure}

Figure \ref{ee-multiplication-circuit} shows an example of a quantum multiplication circuit.  The circuit is composed of (i) conditional addition circuits and  (ii) arrays of Toffoli gates.

As multiplication is a fundamental operation in computation, researchers have developed many circuits to implement several algorithms for the tasks of multiplication such as shift and add, and Karatsuba's algorithm \cite{Kowada} \cite{Edgard2019multiplication}.  However, for quantum computing, only a subset of these algorithms can be used.  Specifically, we cannot use algorithms that result in circuit implementations that depend on feedback or fan out paths.  The multiplication circuit shown in Figure \ref{ee-multiplication-circuit} is based on the shift and add multiplication algorithm and can be viewed as a quantum circuit equivalent to array multipliers such as those presented in \cite{Walter1993arraymultiplier} \cite{Hwang1979Multiplier}.  More formally, the circuit shown in Figure \ref{ee-multiplication-circuit} performs the multiplication of two $n$ bit values $a$ and $b$  by computing the following expression:

\begin{tcolorbox}[fonttitle=\bfseries , colframe=green!90!black, colback=green!10!white]
	\begin{equation}
	P = \sum_{i = 0}^{N-1} a \cdot b_i \cdot 2^i
	\label{encyclopedia-equation-p3}
	\end{equation}
\end{tcolorbox}

\noindent
where $P$ is the product.  

The circuit shown in Figure \ref{ee-multiplication-circuit} takes two $n$ bit inputs $a$ and $b$ (located in quantum registers $\ket{A}$ and $\ket{B}$).  The product $P$ shall appear on $2 \cdot n $ ancillae at the end of computation.  The content in registers $\ket{A}$ and $\ket{B}$ are returned to their original values ($a$ and $b$, respectively) at the end of computation.  As shown in Figure \ref{ee-multiplication-circuit} an additional ancillae is required by the multiplication circuit for a total of $2 \cdot n + 1$ ancillae.  The remaining ancillae is restored to its initial value at the end of computation.  

To calculation of each term of Equation \ref{encyclopedia-equation-p3} is done by either (i) Toffoli gate arrays or (ii) conditional adders.  A conditional adder shall function if the control signal is asserted.  For the example shown in Figure \ref{Conditional-Add-Quantum-ee}, if $\ket{Ctrl} = 1$, the circuit shall compute $\ket{X} + \ket{Y}$, else $\ket{X}$ and $\ket{Y}$ pass through the circuit unchanged.  Therefore, the circuit in Figure \ref{Conditional-Add-Quantum-ee} calculates $\ket{X} + \ket{Ctrl} \cdot \ket{Y}$.  Thus, conditional adders can be used to (i) implement each $a \cdot b_i$ term in equation \ref{encyclopedia-equation-p3} and (ii) perform the summation of these terms indicated in equation \ref{encyclopedia-equation-p3}.  The shifting by $2^i$ (see equation \ref{encyclopedia-equation-p3}) can be accomplished with arrays of CNOT gates (see \cite{Lin}) or via appropriate layout of the conditional adders as is the case in Figure \ref{ee-multiplication-circuit} and works such as \cite{Edgard2019multiplication}.  

By using conditional adders that do not produce garbage outputs (such as the one shown in Figure \ref{Conditional-Add-Quantum-ee}), a multiplication circuit free of garbage outputs can be obtained.  The multiplication circuit shown in Figure \ref{ee-multiplication-circuit} produces no garbage output.  The resource cost of this multiplication circuit depends on the resource usage of its building blocks.  Therefore, by using conditional adders that enjoy low resource costs (such as low T-count), the multiplication circuit as a whole shall benefit.

\subsection{Division}
\label{ee-wiley-division}

\begin{figure}[thbp]
	\flushleft
	\small
	\begin{tcolorbox}[title = \textbf{Algorithm:} Non-Restoring Division Algorithm,fonttitle=\bfseries , colframe=blue!90!black, colback=blue!10!white]
		\begin{tabular}{ll}
			\\ \midrule
			\multicolumn{2}{l}{\textbf{Function} Non-Restoring($a,b$) }\\ \midrule
			\multicolumn{2}{l}{Requirements: $a$ and $b$ are $2$'s complement positive values. }\\
			\multicolumn{2}{l}{\qquad // Takes $a$ and $b$ as two $n$ bit inputs. }\\
			\multicolumn{2}{l}{\qquad // Returns a $n$ bit quotient $Q$ and}\\ 
			\multicolumn{2}{l}{\qquad // a $n-1$ bit remainder $R$. }\\ 
			& \\
			1 & $R = 0$;  \\
			2 & $Q =  a_{n-1}$; // $a_{n-1} \in \{0,1\}$.\\
			3 &\qquad // $a_{n-1}$ is the most significant bit of $a$. \\
			4 & $Q = Q - b$ \\
			5 &\\
			6 & \textbf{For} $i = 1 \text{ to } n-1$ \\ 
			7 & \qquad $Q_{n-i} = \overline{Q_{n-i}}$ \\
			8 & \qquad Concatenate $Q_{n-1-i} \cdots Q_0$ and $R_{n-2} \cdots$  \\
			& \qquad \qquad $\cdots R_{n-1-i}$ then store result in $Y$ \\
			9 & \qquad // $Q_{n-1-i}$ is the most significant bit of $Y$.\\
			10 & \qquad \textbf{If} $(Q_{n-i} = 0)$ \\
			11 & \qquad \qquad $Y = Y + b$ \\
			12 & \qquad \textbf{Else} \\
			13 & \qquad \qquad $Y = Y - b$ \\
			14 & \qquad \textbf{End} \\
			15 & \textbf{End} \\
			16 &\\
			17 & \textbf{If} $(R < 0)$ \\
			18 & \qquad $R = R + b$ \\
			19 & \textbf{End} \\
			20 & $Q_{0} = \overline{Q_{0}}$ \\
			21 &\textbf{Return} $Q, R$ \\ \bottomrule
		\end{tabular}
		
		\caption{ Pseudo-code outline of the non-restoring division algorithm.  }
		\label{sqrt-table:70}
	\end{tcolorbox}
\end{figure}

\begin{figure}[tbhp] 
	\centering
	\begin{tcolorbox}[title = Quantum Division Circuit,fonttitle=\bfseries , colframe=blue!90!black, colback=blue!10!white]
		\[
		\Qcircuit @C = .4em @R = .7em {
			\lstick{\ket{A_0}} & \qw & \qw & \qw &\qw & \qw &  \qw & \multigate{5}{\begin{sideways} Add/Subtract \end{sideways}} & \multigate{3}{\begin{sideways} Ctrl-Add \end{sideways}} & \qw & \qw  & \rstick{\ket{R_0}}\\
			\lstick{\ket{A_1}} & \qw & \qw &\qw & \qw &  \multigate{5}{\begin{sideways} Add/Subtract \end{sideways}} &  \qw & \ghost{\begin{sideways} Add/Subtract \end{sideways}} & \ghost{\begin{sideways} Ctrl-Add \end{sideways}} &  \qw & \qw  &  \rstick{\ket{R_1}} \\
			\lstick{\ket{A_2}} & \qw &\qw &  \multigate{5}{\begin{sideways} Add/Subtract \end{sideways}} & \qw & \ghost{\begin{sideways} Add/Subtract \end{sideways}} &  \qw & \ghost{\begin{sideways} Add/Subtract \end{sideways}} & \ghost{\begin{sideways} Ctrl-Add \end{sideways}} &   \qw & \qw  &  \rstick{\ket{R_2}} \\
			\lstick{\ket{B_{2:0}}} & \multigate{5}{\begin{sideways} Subtract \end{sideways}} & \qw & \ghost{\begin{sideways} Add/Subtract \end{sideways}} & \qw & \ghost{\begin{sideways} Add/Subtract \end{sideways}} &  \qw & \ghost{\begin{sideways} Add/Subtract \end{sideways}} & \ghost{\begin{sideways} Ctrl-Add \end{sideways}} &   \qw & \qw  &  \rstick{\ket{B_{2:0}}} \\
			\lstick{\ket{B_3}} & \ghost{\begin{sideways} Subtract \end{sideways}} &	\qw & \ghost{\begin{sideways} Add/Subtract \end{sideways}} & \qw & \ghost{\begin{sideways} Add/Subtract \end{sideways}} &  \qw & \ghost{\begin{sideways} Add/Subtract \end{sideways}} & \qw &   \qw &  \qw &  \rstick{\ket{B_3}} \\
			\lstick{\ket{A_3}} & \ghost{\begin{sideways} Subtract \end{sideways}} & \qw &\ghost{\begin{sideways} Add/Subtract \end{sideways}} & \qw & \ghost{\begin{sideways} Add/Subtract \end{sideways}} & \qw & \ghost{\begin{sideways} Add/Subtract \end{sideways}} & \ctrl{-2} & \targ & \qw & \rstick{\ket{Q_0}}\\
			\lstick{\ket{0}} & \ghost{\begin{sideways} Subtract \end{sideways}} & \qw & \ghost{\begin{sideways} Add/Subtract \end{sideways}} & \qw & \ghost{\begin{sideways} Add/Subtract \end{sideways}} & \targ & \ctrl{-1} & \qw &  \qw &  \qw & \rstick{\ket{Q_1}} \\
			\lstick{\ket{0}} & \ghost{\begin{sideways} Subtract \end{sideways}} & \qw & \ghost{\begin{sideways} Add/Subtract \end{sideways}} & \targ & \ctrl{-1}& \qw &  \qw &  \qw &  \qw &  \qw & \rstick{\ket{Q_2}} \\
			\lstick{\ket{0}} & \ghost{\begin{sideways} Subtract \end{sideways}} & \targ & \ctrl{-1} & \qw &  \qw &  \qw &  \qw &  \qw &  \qw &  \qw &  \rstick{\ket{Q_3}} \\
		}
		\]	
		\caption{Quantum Circuit for Division.  The circuit shown implements the non-restoring division algorithm.  Source: \cite{Edgard2019divider}}
		\label{quant-division}
	\end{tcolorbox}
\end{figure}
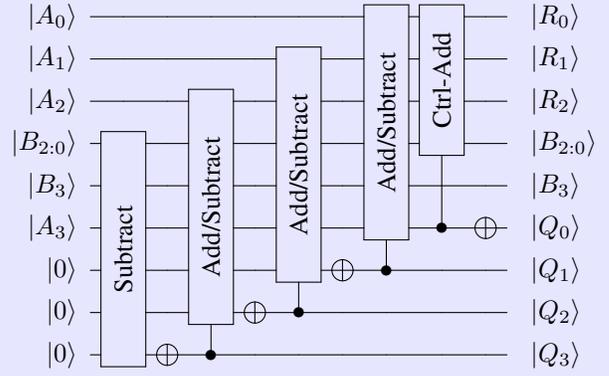

Figure \ref{quant-division} shows an example of a quantum division circuit.  The circuit is composed of (i) a conditional adder, (ii) a subtraction circuit and (iii) a conditional addition and subtraction circuit.  This circuit implements the non-restoring division algorithm.  The non-restoring division algorithm (and its counterpart the non-restoring division algorithm) are well suited to be implemented on a quantum machine.  Other algorithms (such as SRT division or serial implementations) are not suitable for quantum computation due to significant cost overhead.  The building blocks of the division circuit shall be discussed before the operation of the entire division circuit is outlined.

Quantum conditional adders and quantum subtraction circuit were discussed in Section \ref{subtraction-ee-encyclopedua} and Section \ref{addition-ee-encyclopedia}.  A quantum conditional addition and subtraction circuit executes addition or subtraction depending on a control signal.  Figure \ref{subtraction-unit-ee-encyclopedia} presents a generic example of a quantum addition and subtraction circuit.  This example accepts two inputs stored in quantum registers $\ket{A}$ and $\ket{B}$.  The result of computation $Y$ shall appear on register $\ket{B}$ at the end of computation.  An additional qubit ($\ket{Ctrl}$) contains the control signal.  By placing arrays of CNOT gates (which operate on qubit $\ket{Ctrl}$ and the register $\ket{B}$) before and after a quantum addition circuit.  The following computation takes place:

\begin{tcolorbox}[fonttitle=\bfseries , colframe=green!90!black, colback=green!10!white]
	\begin{equation}
	Y = \begin{cases}
	\ket{A} + \ket{B} & \text{ if } \ket{Ctrl} = 0 \\
	\ket{A} - \ket{B} & \text{ if } \ket{Ctrl} = 1 
	\end{cases}
	\label{encyclopedia-equation-p4}
	\end{equation}      
\end{tcolorbox}

Thus, a quantum addition circuit (such as the one shown in Section \ref{addition-ee-encyclopedia}) can be made to perform either addition or subtraction.  

The division circuit shown in Figure \ref{quant-division} takes two $n$ bit numbers $a$ and $b$ that are $2$'s complement positive binary numbers.  $a$ and $b$ and $n-1$ ancillae are applied as inputs to the circuit.  At the end of computation, the quotient of $\frac{a}{b}$, the remainder of the computation of $\frac{a}{b}$ and the input $a$ appear on the circuit outputs.  The quotient $\ket{Q}$ is $n$ bits in size while the remainder $\ket{R}$ is $n-1$ bits in size.  Figure \ref{sqrt-table:70} illustrates how the division circuit presented in this Section executes the division operation.

Reading the circuit from left to right shown in Figure \ref{quant-division}, $Q = Q - b$ (line 4 in the Algorithm shown in Figure \ref{sqrt-table:70}) is calculated first.  The following iterations of conditional addition and subtraction and NOT gates implement each iteration of the \textit{FOR} loop whose contents occupy lines 6 through 15 of the Algorithm (see Figure \ref{sqrt-table:70}).  At the end of each iteration bit $\ket{Q_{n-i}}$ of the quotient is generated (where $1 \leq i \leq n-1$).  The shifting called for in the non-restoring division Algorithm is done without gates in hardware because of the circuit layout (see Figure \ref{quant-division}).  The conditional adder that precedes the production of the remainder $R$ executes the \textit{IF} statement that occupies lines 17 through 19 in the Algorithm in Figure \ref{sqrt-table:70}.  Thus, Figure \ref{quant-division} shows a quantum hardware implementation of the non-restoring division algorithm shown in Figure \ref{sqrt-table:70}.           

The divider shown is free of garbage output as it is based on building blocks which themselves do not generate garbage output (such as those in \cite{Edgard2019divider} \cite{Aghababa2011QuantumDivisder}).  The choice of building blocks plays a large role in determining the resulting resource costs (such as T-count, CNOT-depth, etc.) for the division circuit.  Thus, by selecting addition circuits, subtraction circuits, etc. which enjoy low resource costs for measures of interest, a resource efficient quantum division circuit can be devised.      


\subsection{Quantum Fourier Transform}
\label{QFT}

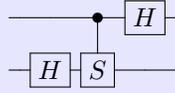
\begin{figure}
	\begin{tcolorbox}[title = Quantum Fourier Transform,fonttitle=\bfseries , colframe=blue!90!black, 	colback=blue!10!white]
		\centering
		\begin{subfigure}[htb]{1.3in}
			\[
			\frac{1}{2} \cdot \begin{bmatrix}
			1 & 1 & 1 & 1 \\
			1 & j & -1 & -j \\
			1 & -1 & 1 & -1 \\
			1 & -j & -1 & j \\
			\end{bmatrix}
			\]
		\end{subfigure} \qquad 	\begin{subfigure}[htb]{1.3in}
			\[
			\Qcircuit @C = .4em @R = .7em {
				&\qw & \qw & \ctrl{1} & \gate{H} & \qw \\
				&\qw & \gate{H} & \gate{S} & \qw & \qw \\
			}
			\]
		\end{subfigure}
		\caption{Implementation of a two qubit quantum Fourier transform.  Matrix representation and circuit diagram are shown.  Hadamard gate and a controlled version of the phase gate are used.}
		\label{a-QFT-build}
	\end{tcolorbox}
\end{figure}

Figure \ref{a-QFT-build} shows a two qubit implementation of the quantum Fourier transform.  The gate schematic and matrix representations are shown.  The computation of the quantum Fourier transform is a manipulation of the superposition of an input quantum state.  We shall illustrate with a two qubit example.  For the case of a two qubit system the quantum Fourier transform equation shown in \ref{encyclopedia-ee-eq7} is rewritten as follows:

 \begin{tcolorbox}[fonttitle=\bfseries , colframe=green!90!black, colback=green!10!white]
 \begin{equation}
 \ket{x} = \frac{1}{2} \sum_{y = 0}^{3} \alpha_y \cdot e^\frac{2 \cdot \pi \cdot j \cdot x \cdot y}{4} \ket{y}
 \label{encyclopedia-ee-eq8}
 \end{equation}
\end{tcolorbox}

where $N = 4$ because two qubits can represent $4$ states and $\alpha_y$ is the probability of the quantum register being in state $y$.  To generalize, for a register of $Z$ qubits, $N = 2^Z$ in Equation \ref{encyclopedia-ee-eq8}.  We can calculate the final superposition state for the quantum register $\ket{X}$ by evaluating Equation \ref{encyclopedia-ee-eq8} for each state value $x$.  At the end of computation, the state probabilities ($\lambda_0, \lambda_1, \lambda_2, \lambda_3$) for the four possible states of $\ket{X}$ ($\ket{00}$, $\ket{01}$, $\ket{10}$ and $\ket{11}$) are given as:

 \begin{tcolorbox}[fonttitle=\bfseries , colframe=green!90!black, colback=green!10!white]
	\begin{equation}
\begin{array}{rcl}
	\lambda_0 \ket{00} & = & \frac{1}{2} \cdot (\alpha_0 + \alpha_1 + \alpha_2 + \alpha_3)\\
	 & & \\
	\lambda_1 \ket{01} & = & \frac{1}{2} \cdot (\alpha_0 + \alpha_1 \cdot e^{\frac{j \cdot \pi}{2}}+ \alpha_2 \cdot e^{j \cdot \pi} \\
		 & & \qquad \qquad + \alpha_3 \cdot e^{\frac{j \cdot 3 \cdot \pi}{2}})\\
		 	 & & \\
 	\lambda_2 \ket{10} & = & \frac{1}{2} \cdot (\alpha_0 + \alpha_1 \cdot e^{j \cdot \pi} + \alpha_2 \cdot e^{j \cdot 2 \cdot \pi} \\
		  & & \qquad \qquad+ \alpha_3 \cdot e^{j \cdot 3 \cdot \pi})\\
		  & & \\
	\lambda_3 \ket{11} & = & \frac{1}{2} \cdot (\alpha_0 + \alpha_1 \cdot e^{\frac{j \cdot 3 \cdot \pi}{2}} + \alpha_2 \cdot e^{j \cdot 3 \cdot \pi} \\
		 & & \qquad \qquad + \alpha_3 \cdot e^{\frac{j \cdot 9 \cdot \pi}{2}})\\	 
\end{array}
 \label{encyclopedia-ee-eq9}
\end{equation}
\end{tcolorbox}
 
 Which can be cleaned up into the following:
 
 \begin{tcolorbox}[fonttitle=\bfseries , colframe=green!90!black, colback=green!10!white]
 	\begin{equation}
 	\begin{array}{rcl}
 	\lambda_0 \ket{00} & = & \frac{1}{2} \cdot (\alpha_0 + \alpha_1 + \alpha_2 + \alpha_3)\\
 	& & \\
 	\lambda_1 \ket{01} & = & \frac{1}{2} \cdot (\alpha_0 +  j \cdot \alpha_1 + -\alpha_2  +  -j \cdot \alpha_3 )\\
 & & \\
 \lambda_2 \ket{10} & = & \frac{1}{2} \cdot (\alpha_0 + -\alpha_1 + \alpha_2 + -\alpha_3)\\
& & \\
\lambda_3 \ket{11} & = & \frac{1}{2} \cdot (\alpha_0 + -j \cdot \alpha_1 + -\alpha_2 + j \cdot \alpha_3)\\	 
\end{array}
\label{encyclopedia-ee-eq10}
\end{equation}
\end{tcolorbox}

As can be seen the calculation has manipulated the superposition state of the quantum register applied to the circuit.  The transformation shown in equations \ref{encyclopedia-ee-eq10} can be represented in matrix form and is shown in Figure \ref{a-QFT-build}.  The matrix transformation can be implemented in terms of quantum gates and the schematic is shown in  Figure \ref{a-QFT-build}.  The circuit is based on the Hadamard gate and a controlled version of the phase gate.  This controlled phase gate is like the CNOT gate in that if the control qubit is one, then the phase gate operation $\begin{bmatrix}
1 & 0 \\
0 & i 
\end{bmatrix}$ is performed on the target qubit.  For additional details on the quantum Fourier transform and additional circuits which operate on quantum registers containing transform domain values, the reader is encouraged to see \cite{Chaung2011QuantumBook} \cite{Lidia2017QFTcircuit} \cite{QFT-circuits}.

\section{Calculation of Quantum Resource Costs}
\label{quantum-costs}

In Section \ref{encycpedia-cost-measure-section} we introduced the cost measures used to assess the resource costs of quantum circuits for the cases of (i) fault tolerant quantum computers and (ii) noisy intermediate scale quantum (NISQ) computers.  Having now seen several quantum circuits for elementary arithmetic operations, we can now show how to compute resource costs for these circuits.  The purpose of this Section is to illustrate how one goes about calculating the resource costs of a quantum circuit.  The computation of resource costs of fault tolerant circuits and the computation of resource costs of circuits for NISQ machine shall be handled separately

\subsection{Resource Cost Calculations for a Circuit for a Fault Tolerant Quantum Computer}

For this example, we shall seek to calculate the T-count, T-depth, garbage output, qubit cost and $KQ_T$ for the quantum addition circuit presented in Section \ref{addition-ee-encyclopedia} and shown in Figure \ref{adder-unit}.  The measures T-count, T-depth, qubit cost and $KQ_T$ have been introduced in Section \ref{encycpedia-cost-measure-section}.  The quantum adder is composed of functional blocks (called Unmajority and Add ($UMA$) and Majority ($MAJ$)) which in turn are constructed of Toffoli and CNOT gates (see Figure \ref{adder-unit}).  Once a quantum circuit has been decomposed in terms of logic gates (i.e. Toffoli and CNOT gates), we can proceed to determine the cost measures.  The CNOT gate is already a member of the Clifford+T gate family (see Figure \ref{Clifford table}) while the Toffoli gate needs to be decomposed into a Clifford+T gate implementation  We shall use the implementation shown in Figure \ref{overview:clifTtoffoli} and presented in \cite{Maslov}.  Having decomposed the circuit from a top level description to its Clifford+T gate circuitry, we can follow the procedure below to compute the T-depth and T-count.

\begin{figure}[thbp]
	\begin{tcolorbox}[title = Procedure to calculate T-Count and T-Depth,fonttitle=\bfseries , colframe=blue!90!black, colback=blue!10!white]
		\begin{itemize}
			\item \textbf{Step 1:} Decompose the top-level circuit to its equivalent Clifford+T gate networks.  As needed decompose circuit to submodules or logic gates to simplify the decomposition.
			\item \textbf{Step 2:} Calculate the T-count and T-depth for any submodules or logic gates from Step 1.   
			\item \textbf{Step 3:} Calculate the T-count and T-depth for the whole circuit using results from Step 2.
		\end{itemize}
	\end{tcolorbox}
\end{figure}

Thus, we shall now calculate the T-count and T-depth of the Toffoli gate.  The Clifford+T gate implementation of the Toffoli gate is repeated in Figure \ref{overview:clifTtoffoli}. 

\begin{figure}[htbp]
	\centering
	\begin{tcolorbox}[title = T-Depth of the Toffoli Gate,fonttitle=\bfseries , colframe=blue!90!black, colback=blue!10!white]
		\includegraphics[width = 3.15in]{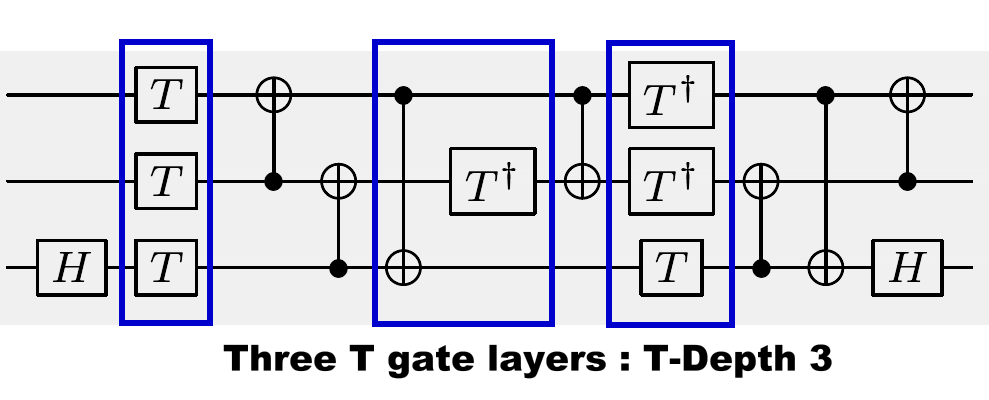}
		\caption{Clifford+T gate implementation of the Toffoli gate illustrating the T-depth (and the corresponding gate layers).  Implementation presented in \cite{Maslov}.}
		\label{clifT-Tdepth}
	\end{tcolorbox}	
\end{figure}

To calculate T-count, we count the number of T gates in the Toffoli gate circuit.  As shown in Figure \ref{clifT-Tdepth}, the T-count is $7$.  To calculate T-depth, we shall follow the procedure presented in \cite{Selinger2013Tgatelayers} \cite{Maslov} \cite{Niemann2019Tgatelayers}.  In \cite{Selinger2013Tgatelayers} \cite{Maslov} \cite{Niemann2019Tgatelayers}, the circuit is viewed as a series of gate layers (where a gate layer consists of quantum gates that run in parallel).  Thus, the T-depth is the number of gate layers which contain 1 or more T gates.  For the Toffoli gate in Figure \ref{clifT-Tdepth}, the T-depth is $3$.  

We can now consider the costs of the $MAJ$ and $UMA$ functional blocks (see Figure \ref{adder-unit}).  As shown in Figure \ref{adder-unit}, both of these functional units contain 1 Toffoli gate apiece and the remaining gates are CNOT gates.  For both the $MAJ$ and $UMA$ functional blocks, the T-count is $7$ and the T-depth is $3$.  Now we can turn our attention to the adder itself.  As shown in Figure \ref{adder-unit}, the quantum adder consists of $n$ $MAJ$ units in series (on a diagonal), a CNOT gate and finally $n$ $UMA$ units in series (again on a diagonal). The T-count is given as the total of $MAJ$ and $UMA$ functional blocks multiplied by the T-count per functional block.  The mathematical computation is as follows:

\begin{tcolorbox}[fonttitle=\bfseries , colframe=green!90!black, colback=green!10!white]
	\begin{equation}
	7 \cdot (2 \cdot n) = 14 \cdot n.     
	\label{encyclopedia-equation-p15}
	\end{equation}
\end{tcolorbox}

Since the $MAJ$ and $UMA$ functional blocks are in series, these modules form $2 \cdot n$ layers.  Each layer contains a Toffoli gate with an associated T-depth of $3$.  Therefore, for this circuit the T-depth is given as the total of $MAJ$ and $UMA$ functional blocks multiplied by the T-depth per functional block.  The mathematical computation is as follows:

\begin{tcolorbox}[fonttitle=\bfseries , colframe=green!90!black, colback=green!10!white]
	\begin{equation}
	3 \cdot (2 \cdot n) = 6 \cdot n.     
	\label{encyclopedia-equation-p14}
	\end{equation}
\end{tcolorbox}

We still need to compute the garbage output cost, qubit cost and $KQ_T$ metrics for the quantum circuit.  We shall calculate qubit cost first because with the qubit cost, we shall be equipped to calculate the remaining measures.  Since the number of qubits required by a quantum circuit remains constant, we have two methods at our disposal to compute the qubit cost:

\begin{tcolorbox}[title = Methods to Calculate Qubit Count,fonttitle=\bfseries , colframe=blue!90!black, colback=blue!10!white]
	\begin{itemize}
		\item \textbf{Method 1: } Qubit cost = ancillae + primary inputs
		\item \textbf{Method 2: } Qubit cost = garbage output + circuit outputs (where circuit outputs include primary inputs and restored ancillae)
	\end{itemize}
\end{tcolorbox}

We shall use \textit{Method 1}.  The quantum adder takes two $n$ bit inputs and requires two ancillae (see Figure \ref{adder-unit} and Section \ref{addition-ee-encyclopedia}).  Hence the qubit cost is given as:

\begin{tcolorbox}[fonttitle=\bfseries , colframe=green!90!black, colback=green!10!white]
	\begin{equation}
	2 \cdot n + 2
	\label{encyclopedia-equation-p16}
	\end{equation}
\end{tcolorbox}

Now we can turn our attention to the garbage output cost.  At the end of computation, the quantum adder produced a $n+1$ sum value, the remaining ancilla is restored to $0$ and one of the $n$ bit inputs is restored (see Section \ref{addition-ee-encyclopedia}).  To determine the garbage output, we solve the \textit{Method 2} expression of qubit cost for the garbage output term.  As a result, we obtain the following calculation:

\begin{tcolorbox}[fonttitle=\bfseries , colframe=green!90!black, colback=green!10!white]
	\begin{equation}
	(2 \cdot n + 2) - (n + 1) - n - 1 \rightarrow 0
	\label{encyclopedia-equation-p17}
	\end{equation}
\end{tcolorbox}

As shown in Equation \ref{encyclopedia-equation-p17}, the addition circuit is free of garbage output.  The $KQ_T$ measure is calculated as the product of the T-depth $(K)$ and qubit cost $(Q)$.  The qubit cost is in Equation \ref{encyclopedia-equation-p16} and T-depth is in Equation \ref{encyclopedia-equation-p14}.  Thus the $KQ_T$ measure is given as:

\begin{tcolorbox}[fonttitle=\bfseries , colframe=green!90!black, colback=green!10!white]
	\begin{equation}
	(6 \cdot n) \cdot (2 \cdot n + 2) = 12 \cdot n^2 + 12 \cdot n
	\label{encyclopedia-equation-p18}
	\end{equation}
\end{tcolorbox}

\subsection{Resource Cost Calculations for a Circuit for Noisy Intermediate Scale Quantum Computers}

For this example, we now consider the cost measures for quantum circuits that shall run on NISQ machines.  Thus, we shall illustrate the calculation of the CNOT-count, CNOT-depth, qubit cost, garbage output and $KQ_{CNOT}$.  These measures were presented in Section \ref{encycpedia-cost-measure-section} and in Figure \ref{NISSQ-cost-measures}.  We shall consider the conditional adder presented in Figure \ref{Conditional-Add-Quantum-ee} in Section \ref{multiplication-section} for our computations.  We begin by decomposing the circuit.  The conditional addition circuit is composed of fundamental building blocks (labeled $CUMA$ and $CMAJ$).  These functional blocks in turn can be decomposed to Toffoli and CNOT gates.  We note that the CNOT gate is already a Clifford+T gate and we shall use the Clifford+T gate implementation of the Toffoli gate shown in Figure \ref{overview:clifTtoffoli} and presented in \cite{Maslov}.  We are now in a position to calculate CNOT-count and CNOT-depth and shall use a procedure to calculate these measures with is identical to the example for fault tolerant quantum machines:

\begin{figure}[thbp]
	\begin{tcolorbox}[title = Procedure to calculate CNOT-Count and CNOT-Depth,fonttitle=\bfseries , colframe=blue!90!black, colback=blue!10!white]
		\begin{itemize}
			\item \textbf{Step 1:} Decompose the top-level circuit to its equivalent Clifford+T gate networks.  As needed decompose circuit to submodules or logic gates to simplify the decomposition.
			\item \textbf{Step 2:} Calculate the CNOT-count and CNOT-depth for any submodules or logic gates from Step 1.   
			\item \textbf{Step 3:} Calculate the CNOT-count and CNOT-depth for the whole circuit using results from Step 2.
		\end{itemize}
	\end{tcolorbox}
\end{figure}

\begin{figure}[hbtp]
	\centering
	\begin{tcolorbox}[title = CNOT-Depth of the Toffoli gate,fonttitle=\bfseries , colframe=blue!90!black, colback=blue!10!white]
		\includegraphics[width = 3.15in]{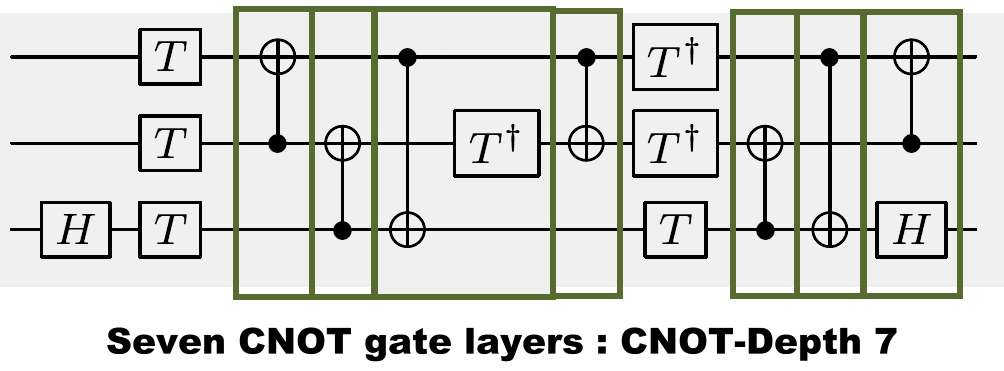}
		\caption{Illustration of the Clifford+T implementation of the Toffoli gate.  The gate layers with CNOT gates are shown.}
		\label{toffoli-CNOTdepth}
	\end{tcolorbox}
\end{figure}

We shall begin with the Toffoli gate implementation shown in Figure \ref{toffoli-CNOTdepth}.  To determine CNOT-count, we count up the CNOT gates used in the circuit.  For the example shown in Figure \ref{toffoli-CNOTdepth}, the CNOT-count is $7$.  To calculate CNOT-depth, we again view the circuit as a series of gate layers.  As shown in Figure \ref{toffoli-CNOTdepth}, we see that seven gate layers contain CNOT gates.  Thus, CNOT-depth is $7$.  With the CNOT-count and CNOT-depth of the Toffoli gate found, we can now proceed to larger building blocks.  The conditional adder is composed of Controlled Majority ($CMAJ$) and Controlled Unmajority and Add ($CUMA$) functional blocks (see Figure \ref{Conditional-Add-Quantum-ee}).  The $CMAJ$ building block consists of $2$ Toffoli gates and a CNOT gate.  The logic gates are in series.  Thus, the CNOT-count is $ 7 \cdot 2 + 1 = 15 $ and the CNOT-depth is $ 7 \cdot 2 + 1 = 15 $.  Likewise, the $CUMA$ also is constructed from $2$ Toffoli gates and a CNOT gate.  These are also in series.  As a result the CNOT-count is $15$ and the CNOT-depth is $15$.  

Now we can turn our attention to the quantum conditional adder itself.  The quantum conditional adder consists of $n$ $CMAJ$ units in series (on a diagonal), a Toffoli gate, then $n$ $CUMA$ units in series (again on a diagonal).  All gates are in series meaning each building block listed forms a gate layer.  Thus, the CNOT-count is given as the total of the Toffoli gate, $CMAJ$ and $CUMA$ functional blocks multiplied by the CNOT-count per functional block.  Each $CUMA$ and $CMAJ$ module has a CNOT-count of $15$ each and the Toffoli gate has a CNOT-count of $7$.  The computation is as follows:

\begin{tcolorbox}[fonttitle=\bfseries , colframe=green!90!black, colback=green!10!white]
	\begin{equation}
	15 \cdot n + 15 \cdot n + 7 \rightarrow 30 \cdot n + 7
	\label{encyclopedia-equation-p19}
	\end{equation}
\end{tcolorbox}

To calculate CNOT-depth, note that the CMAJ blocks, Toffoli gate and CUMA blocks are in series.  Thus, the total CNOT-depth is the sum total of the CNOT-depths of these functional blocks.  Each $CUMA$ and $CMAJ$ module has a CNOT-depth of $15$ each and the Toffoli gate has a CNOT-depth of $7$.  The computation of total CNOT-depth is as follows:

\begin{tcolorbox}[fonttitle=\bfseries , colframe=green!90!black, colback=green!10!white]
	\begin{equation}
	15 \cdot n + 15 \cdot n + 7 \rightarrow 30 \cdot n + 7
	\label{encyclopedia-equation-p20}
	\end{equation}
\end{tcolorbox}

Now we can turn our attention to the qubit cost, garbage output cost and $KQ_{CNOT}$ measures.  We shall first calculate the qubit cost.  We calculate the qubit cost by finding the sum of the two $n$ bit inputs and three ancillae.  Thus, we shall use \textit{Method 1} (Qubit cost = ancillae + primary inputs).  The expression for the qubit cost is given as:

\begin{tcolorbox}[fonttitle=\bfseries , colframe=green!90!black, colback=green!10!white]
	\begin{equation}
	2 \cdot n + 3 
	\label{encyclopedia-equation-p21}
	\end{equation}
\end{tcolorbox}

To calculate garbage output, we take the total qubit cost and subtract $n+1$ qubits occupied by the output, $n$ qubits holding one of the original operands, the qubit $\ket{Ctrl}$ and all ancillae that are restored to their initial values.  Thus, the garbage output is calculated as:

\begin{tcolorbox}[fonttitle=\bfseries , colframe=green!90!black, colback=green!10!white]
	\begin{equation}
	(2 \cdot n + 3) - (n + 1) - (n+1) - 1 \rightarrow 0
	\label{encyclopedia-equation-p12}
	\end{equation}
\end{tcolorbox}

Equation \ref{encyclopedia-equation-p12} indicates that the conditional adder produces no garbage output.  The $KQ_{CNOT}$ measure is the product of qubit cost $Q$ and the CNOT-depth $K$.  Having already calculated qubit cost and CNOT-depth we can present the $KQ_{CNOT}$ as:

\begin{tcolorbox}[fonttitle=\bfseries , colframe=green!90!black, colback=green!10!white]
	\begin{equation}
	(30 \cdot n + 7) \cdot (2 \cdot n + 3) = 60 \cdot n^2 + 104 \cdot n + 21
	\label{encyclopedia-equation-p13}
	\end{equation}
\end{tcolorbox}

\section{Application of Quantum Arithmetic Circuits in Quantum Algorithms}
\label{example-circuit-implementation}

The arithmetic units presented in this work can be combined into more complex functional blocks for use in quantum algorithms. We present an example where quantum circuits for addition and multiplication are arranged into a functional block that performs image rotation.  The design presented is shown in \cite{FeiYan2017QuantumRotation}.  We present a quantum implementation of image rotation because quantum computing for image processing applications has caught the attention of researchers.  The availability of quantum circuits for image processing applications can benefit fields such as medicine and scientific computation (see \cite{YanFei2020CriticalQuantumImageProcessing} \cite{Iliyasu2013useofrotation} \cite{Lehmann1999useofrotation}  \cite{Haponen2003useofimagerotation} \cite{Huifang2015useofrotation}).  As a result, quantum algorithms for image processing have been developed as shown in \cite{ChengZhenwen2018QuantumWatermarking} \cite{YanFei2020CriticalQuantumImageProcessing} \cite{Yu2019QuantumImageTracking} \cite{Song2014QuantumImageEnceryption}.  Quantum circuits for operations such as image rotation are needed to (i) realize quantum algorithms for image processing or (ii) to implement hardware for additional useful tasks such as image registration and image fusion \cite{FeiYan2017QuantumRotation}.

In this Section, we present a quantum circuit for the computation of image rotation that is presented in \cite{FeiYan2017QuantumRotation}.   The circuit is designed to operate on image information encoded on qubits using the Novel Enhanced Quantum Representation (NEQR) \cite{Zhang2013quantumimagerepresentation} \cite{FeiYan2017QuantumRotation}.  Details on NEQR can be seen in \cite{Zhang2013quantumimagerepresentation}.  The rotation operation only works with the pixel location information of the image.  In NEQR, the image location information is stored in two quantum registers (see \cite{Zhang2013quantumimagerepresentation}).  The presented circuit uses quantum circuits for addition, subtraction and multiplication (such as those presented in this work) as building blocks.

Given an image pixel with coordinates $x_0, y_0$, the quantum image rotation circuit performs the following computations:

\begin{tcolorbox}[fonttitle=\bfseries , colframe=green!90!black, colback=green!10!white]
	\begin{equation}
	\begin{array}{rcl}
	y_t & = & cos(\theta) \cdot y_0 + sin(\theta) \cdot x_0 \\
	x_t & = & -sin(\theta) \cdot y_0 + cos(\theta) \cdot x_0 
	\end{array}
	\label{encyclopedia-equation-p5}
	\end{equation}
\end{tcolorbox}

\noindent
where $\theta$ is the rotation angle and the sign indicates the direction of rotation ($-$ counter-clockwise and $+$ clockwise) \cite{FeiYan2017QuantumRotation}.  Equation \ref{encyclopedia-equation-p6} can be presented in terms of matrices as:

\begin{tcolorbox}[fonttitle=\bfseries , colframe=green!90!black, colback=green!10!white]
	\begin{equation}
	\begin{bmatrix}
	y_t \\
	x_t \end{bmatrix} = \begin{bmatrix} 
	cos(\theta) &  sin(\theta) \\
	-sin(\theta) & cos(\theta) \end{bmatrix} \cdot \begin{bmatrix}
	y_0 \\
	x_0 \end{bmatrix}
	\label{encyclopedia-equation-p6}
	\end{equation}
\end{tcolorbox}

\begin{figure}
	\centering
	\begin{tcolorbox}[title = Top Level View of Quantum Image Rotation Circuit,fonttitle=\bfseries , colframe=blue!90!black, colback=blue!10!white] 
		\begin{subfigure}[tbhp]{3in}
			\[
			\Qcircuit @C = .4em @R = .7em {
				\lstick{\ket{x_{n-1:0}}} &  \multigate{5}{\begin{sideways} X-axis shear \end{sideways}} & \qw & \multigate{5}{\begin{sideways} Y-axis shear \end{sideways}} & \qw & \multigate{5}{\begin{sideways} X-axis shear \end{sideways}} & \qw & \rstick{\ket{x_{n-1:0}}} \\
				\lstick{\ket{y_{n-1:0}}} & \ghost{\begin{sideways} X-axis shear \end{sideways}} & \qw & \ghost{\begin{sideways} Y-axis shear \end{sideways}} & \qw & \ghost{\begin{sideways} X-axis shear \end{sideways}} & \qw & \rstick{\ket{y_{n-1:0}}} \\
				\lstick{\ket{\alpha}} & \ghost{\begin{sideways} X-axis shear \end{sideways}} & \qw & \ghost{\begin{sideways} Y-axis shear \end{sideways}} & \qw & \ghost{\begin{sideways} X-axis shear \end{sideways}} & \qw & \rstick{\ket{\alpha}} \\
				\lstick{\ket{\beta}} & \ghost{\begin{sideways} X-axis shear \end{sideways}} & \qw & \ghost{\begin{sideways} Y-axis shear \end{sideways}} & \qw & \ghost{\begin{sideways} X-axis shear \end{sideways}} & \qw & \rstick{\ket{\beta}} \\
				\lstick{\ket{X_{ref}}} & \ghost{\begin{sideways} X-axis shear \end{sideways}} & \qw & \ghost{\begin{sideways} Y-axis shear \end{sideways}} & \qw & \ghost{\begin{sideways} X-axis shear \end{sideways}} & \qw & \rstick{\ket{x_s}} \\ 
				\lstick{\ket{Y_{ref}}} &  \ghost{\begin{sideways} X-axis shear \end{sideways}} & \qw & \ghost{\begin{sideways} Y-axis shear \end{sideways}} & \qw & \ghost{\begin{sideways} X-axis
					shear \end{sideways}} & \qw & \rstick{\ket{y_s}} \\
			}
			\]
			\caption{Top level view of the quantum circuit for the rotation operation.}
		\end{subfigure}
		\\ \vspace{.25in}
		\begin{subfigure}[tbhp]{3in}
			\includegraphics[width = 3in]{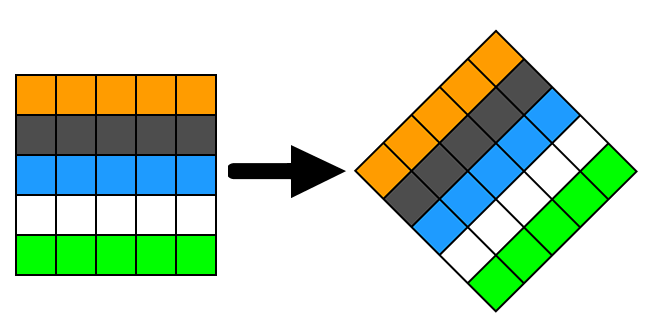}
			\caption{Visualization of the image rotation operation.}
		\end{subfigure}
		\caption{ Quantum Circuit for the Rotation of an NEQR encoded image.  Top level circuit and example image rotation is shown.}
		\label{QIP-top-level}
	\end{tcolorbox}
\end{figure}

The design methodology used in \cite{FeiYan2017QuantumRotation} to implement equation \ref{encyclopedia-equation-p6} splits the rotation task into three separate shearing operations.  Figure \ref{QIP-top-level} shows a top level implementation of the rotation circuit where $x_s, y_s$ are the final image pixel location coordinates.  Also, $tan(\frac{\theta}{2}) = \alpha$, $sin(\theta) = \beta$, $X_{ref}$ and $Y_{ref}$ are constants which corresponds to the center of rotation.   By splitting the rotation into three shearing operations, equation \ref{encyclopedia-equation-p6} can be rewritten in terms of three shearing operations as now shown:

\begin{tcolorbox}[fonttitle=\bfseries , colframe=green!90!black, colback=green!10!white]
	\begin{equation}
	\begin{bmatrix}
	y_t \\
	x_t \end{bmatrix} = \begin{bmatrix} 
	1 &  0 \\
	tan(\frac{\theta}{2}) & 1 \end{bmatrix} \cdot \begin{bmatrix} 
	1 & -sin(\theta)   \\
	0  &  1 \end{bmatrix} \cdot \begin{bmatrix} 
	1 & 0  \\
	tan(\frac{\theta}{2}) & 1 \end{bmatrix} \cdot \begin{bmatrix}
	y_0 \\
	x_0 \\ \end{bmatrix}
	\label{encyclopedia-equation-p7}
	\end{equation}
\end{tcolorbox}

\noindent
where $\theta$ is the rotation angle.  The transformation matrix $\begin{bmatrix} 
1 &  0 \\
tan(\frac{\theta}{2}) & 1 \end{bmatrix}$ represents the following horizontal shear translation:

\begin{tcolorbox}[fonttitle=\bfseries , colframe=green!90!black, colback=green!10!white]
	\begin{equation}
	\begin{array}{lcr}
	(x,y) & = & (x + y \cdot tan \left(\frac{\theta}{2} \right),y) \\
	\end{array}
	\label{encyclopedia-equation-p8}
	\end{equation}
\end{tcolorbox}

The transformation matrix $\begin{bmatrix} 
1 &  -sin{\theta} \\
0 & 1 \end{bmatrix}$ represents the following vertical shear translation:

\begin{tcolorbox}[fonttitle=\bfseries , colframe=green!90!black, colback=green!10!white]
	\begin{equation}
	\begin{array}{lcr}
	(x,y) & = & (x,y - x \cdot sin \left(\frac{\theta}{2} \right)) \\
	\end{array}
	\label{encyclopedia-equation-p9}
	\end{equation}
\end{tcolorbox}

The choice of the centroid of rotation $X_{ref}$  or $Y_{ref}$ is a design choice.  In  \cite{FeiYan2017QuantumRotation}, the centroid corresponds to the center pixel location values of the image.  For the case of image centroids ($X_{ref}$  or $Y_{ref}$) within the range of pixel location values, the final image shearing in the vertical direction becomes:

\begin{tcolorbox}[fonttitle=\bfseries , colframe=green!90!black, colback=green!10!white]
	\begin{equation}
	(x,y) = \begin{cases}
	(x, y + (X_{ref} - x) \cdot sin(\theta)) \text{ if } y \in [y_0, y_i] \\
	(x, y - (x - X_{ref}) \cdot sin(\theta)) \\
	\qquad \qquad \qquad \text{ if } y \in [y_{i+1}: y_{n-1}] 
	\end{cases}
	\label{encyclopedia-equation-p10}
	\end{equation}			
\end{tcolorbox}

The final image shearing in the horizontal direction becomes:

\begin{tcolorbox}[fonttitle=\bfseries , colframe=green!90!black, colback=green!10!white]
	\begin{equation}
	(x,y) = \begin{cases}
	(x - (Y_{ref} - y) \cdot tan \left(\frac{\theta}{2} \right),y ) \text{ if } x \in [x_0, x_i] \\ 
	(x + (y - Y_{ref}) \cdot tan \left(\frac{\theta}{2} \right),y ) \\
	\qquad \qquad \qquad \text{ if } x \in [x_{i+1}, x_{n-1}]
	\end{cases}
	\label{encyclopedia-equation-p11}
	\end{equation}			
\end{tcolorbox}

We require quantum arithmetic circuits for multiplication addition and subtraction.  For each expression shown in Equations \ref{encyclopedia-equation-p10} and \ref{encyclopedia-equation-p11}, an independent circuit is created.

\subsection{Quantum Circuits for Horizontal and Vertical Shear Operations}

\begin{figure}
	\centering
	\begin{tcolorbox}[title = Shear Operation Visualization,fonttitle=\bfseries , colframe=blue!90!black, colback=blue!10!white]
		
		\begin{subfigure}[tbhp]{3in}
			\centering
			\includegraphics[width = 3in]{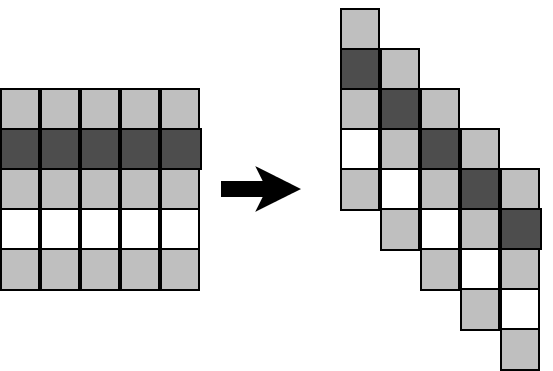}
			\caption{Example of the vertical shear operation performed by the quantum circuit shown in this graphic.}
			\label{image-example-graphics}
		\end{subfigure} 
		\\ 
		\begin{subfigure}[tbhp]{3in}
			\centering
			\includegraphics[width = 3in]{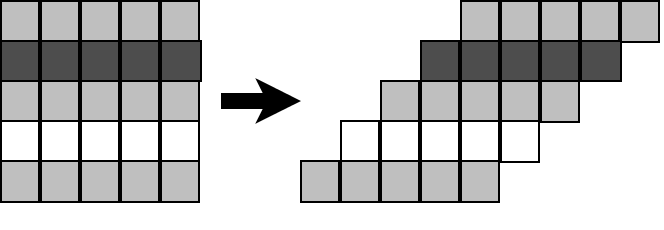}
			\caption{Example of the horizontal shear operation performed by the quantum circuit shown in this graphic.}
			\label{image-example-graphics-2}
		\end{subfigure}
		\caption{Visualizations of the shearing operations.  Horizontal and vertical shearing are shown.} 
	\end{tcolorbox}
\end{figure}

\begin{figure}
	\centering
	\begin{tcolorbox}[title = Vertical Shear Circuit for Pixels Located at $y > Y_{ref}$ where $Y_{ref}$ is the Centroid of Rotation,fonttitle=\bfseries , colframe=blue!90!black, colback=blue!10!white]
		
		\begin{subfigure}[tbhp]{3in}
			\[
			\Qcircuit @C = .4em @R = .7em @!R{
				\lstick{\ket{x_{[n-1:0]}}} & \qw & 	\qw & \qw & \rstick{\ket{x_{[n-1:0]}}} \\
				\lstick{\ket{\alpha_{[n-1:0]}}} & \qw & \qw &  \qw & \rstick{\ket{\alpha_{[n-1:0]}}} \\
				\lstick{\ket{y_{n-1}}} &\ctrl{3} & \ctrl{1} & \qw & \rstick{\ket{y_{n-1}}} \\
				\lstick{\ket{y_{[n-2:0]}}} &\qw & \ctrl{1} & \qw & \rstick{\ket{y_{[n-2:0]}}} \\
				\lstick{\ket{Y_{ref}[n-1:0]}} &\qw & \gate{-} & \qw & \\
				\lstick{\ket{0}} & \targ & \qw &  \qw & \rstick{\ket{y_{n-1}}} 
			}
			\]    
			\caption{Circuit of the vertical shear: Step 1.}
			\label{v-shear-step-1}	
		\end{subfigure} 
		\\
		\begin{subfigure}[tbhp]{3in}
			\[
			\Qcircuit @C = .4em @R = .7em @!R{
				\lstick{\ket{x_{[n-1:0]}}} & \qw & 	\qw & \qw & \qw & \rstick{\ket{x_{[n-1:0]}}} \\
				\lstick{\ket{\alpha_{[n-1:0]}}} & \qw & \qw & \ctrl{3} & \qw & \rstick{\ket{\alpha_{[n-1:0]}}} \\
				\lstick{\ket{y_{n-1}}} &\ctrl{3} & \ctrl{1} & \qw & \qw &   \rstick{\ket{y_{n-1}}} \\
				\lstick{\ket{y_{[n-2:0]}}} &\qw & \ctrl{1} & \qw & \qw &   \rstick{\ket{y_{[n-2:0]}}} \\
				\lstick{\ket{Y_{ref}[n-1:0]}} &\qw & \gate{-} & \gate{\times} & \qw & \\
				\lstick{\ket{0}} & \targ & \qw & \ctrl{-1} & \qw & \rstick{\ket{y_{n-1}}} 
			}
			\]    
			\caption{Circuit of the vertical shear: Step 2.}
			\label{v-shear-step-2} 
			
		\end{subfigure} 
		\\
		\begin{subfigure}[tbhp]{3in} 
			\[
			\Qcircuit @C = .4em @R = .7em @!R{
				\lstick{\ket{x_{[n-1:0]}}} & \qw & 	\qw & \qw & \qw & \qw & \rstick{\ket{x_{[n-1:0]}}} \\
				\lstick{\ket{\alpha_{[n-1:0]}}} & \qw & \qw & \ctrl{3} & \qw &  \qw & \rstick{\ket{\alpha_{[n-1:0]}}} \\
				\lstick{\ket{y_{n-1}}} &\ctrl{3} & \ctrl{1} & \qw & \qw &  \qw & \rstick{\ket{y_{n-1}}} \\
				\lstick{\ket{y_{[n-2:0]}}} &\qw & \ctrl{1} & \qw & \qw &  \qw &  \qw & \rstick{\ket{y_{[n-2:0]}}} \\
				\lstick{\ket{Y_{ref}[n-1:0]}} &\qw & \gate{-} & \gate{\times} & \gate{IP} & \qw &  \\
				\lstick{\ket{0}} & \targ & \qw & \ctrl{-1} & \qw &  \qw & \rstick{\ket{y_{n-1}}} 
			}
			\]    
			\caption{Circuit of the vertical shear: Step 3.}
			\label{v-shear-step-3} 
			
		\end{subfigure} 
		\\
		\begin{subfigure}[tbhp]{3in}
			\[
			\Qcircuit @C = .4em @R = .7em @!R{
				\lstick{\ket{x_{[n-1:0]}}} & \qw & 	\qw & \qw & \qw & \ctrl{4} & \qw & \rstick{\ket{x_{[n-1:0]}}} \\
				\lstick{\ket{\alpha_{[n-1:0]}}} & \qw & \qw & \ctrl{3} & \qw &  \qw &  \qw & \rstick{\ket{\alpha_{[n-1:0]}}} \\
				\lstick{\ket{y_{n-1}}} &\ctrl{3} & \ctrl{1} & \qw & \qw &  \qw &  \qw & \rstick{\ket{y_{n-1}}} \\
				\lstick{\ket{y_{[n-2:0]}}} &\qw & \ctrl{1} & \qw & \qw &  \qw &  \qw & \rstick{\ket{y_{[n-2:0]}}} \\
				\lstick{\ket{Y_{ref}[n-1:0]}} &\qw & \gate{-} & \gate{\times} & \gate{IP} & \gate{+} & \qw & \rstick{\ket{y_s}} \\
				\lstick{\ket{0}} & \targ & \qw & \ctrl{-1} & \qw &  \qw &  \qw & \rstick{\ket{y_{n-1}}} 
			}
			\]    
			\caption{Circuit of the vertical shear: Step 4..  $y_s$ is the result of computation.}
			\label{v-shear-step-4} 
			
		\end{subfigure} 
		\caption{Quantum circuit for the vertical shear of images for pixel located at values greater then the centroid value.  Adapted from \cite{FeiYan2017QuantumRotation}.}
		\label{top-v-part-image}
	\end{tcolorbox}
\end{figure}
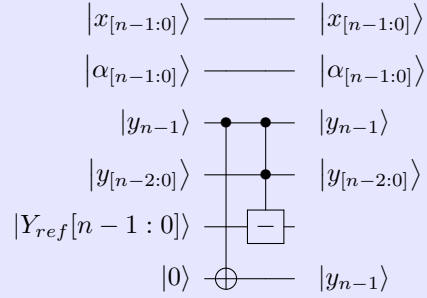
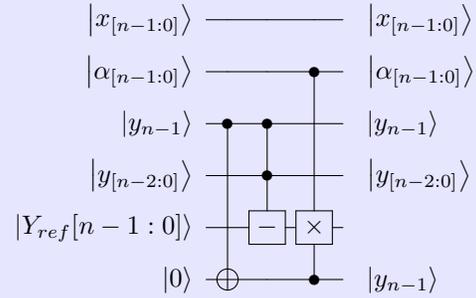
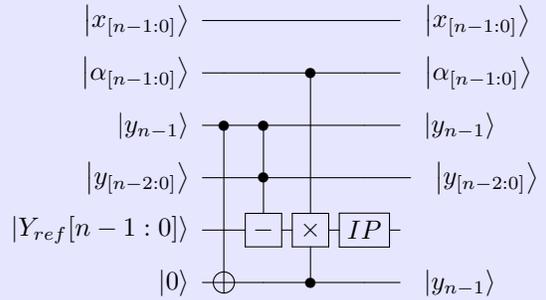
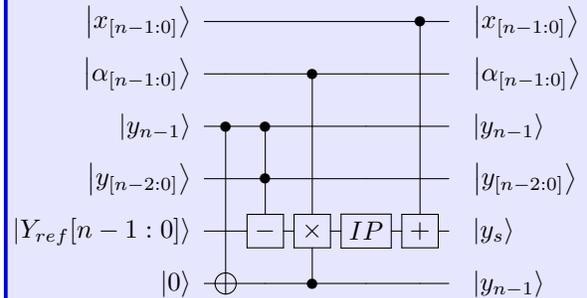

\begin{figure}
	\centering
	\begin{tcolorbox}[title = Horizontal Shear Circuit for Pixels Located at $x > X_{ref}$ where $X_{ref}$ is the Centroid of Rotation,fonttitle=\bfseries , colframe=blue!90!black, colback=blue!10!white]
		\begin{subfigure}[tbhp]{3in}
			\[
			\Qcircuit @C = .4em @R = .7em @!R{
				\lstick{\ket{y_{[n-1:0]}}} & \qw & 	\qw & \qw & \rstick{\ket{y_{[n-1:0]}}} \\
				\lstick{\ket{\beta_{[n-1:0]}}} & \qw & \qw &  \qw & \rstick{\ket{\beta_{[n-1:0]}}} \\
				\lstick{\ket{x_{n-1}}} &\ctrl{3} &  \ctrl{2} & \qw & \rstick{\ket{x_{n-1}}} \\
				\lstick{\ket{x_{[n-2:0]}}} &\qw & \ctrl{1} & \qw & \rstick{\ket{x_{[n-2:0]}}} \\
				\lstick{\ket{X_{ref}[n-1:0]}} &\qw & \gate{-} & \qw & \\
				\lstick{\ket{0}} & \targ & \qw & \qw & \rstick{\ket{x_{n-1}}} 
			}
			\]    
			\caption{Circuit of the horizontal shear: Step 1.}
			\label{h-shear-step-1} 
		\end{subfigure} 
		\\
		\begin{subfigure}[tbhp]{3in}
			\[
			\Qcircuit @C = .4em @R = .7em @!R{
				\lstick{\ket{y_{[n-1:0]}}} & \qw & 	\qw & \qw & \qw & \qw & \rstick{\ket{y_{[n-1:0]}}} \\
				\lstick{\ket{\beta_{[n-1:0]}}} & \qw & \qw & \ctrl{3} & \qw &  \qw & \rstick{\ket{\beta_{[n-1:0]}}} \\
				\lstick{\ket{x_{n-1}}} &\ctrl{3} &  \ctrl{2} & \qw & \qw &  \qw & \rstick{\ket{x_{n-1}}} \\
				\lstick{\ket{x_{[n-2:0]}}} &\qw & \ctrl{1} & \qw & \qw &  \qw & \rstick{\ket{x_{[n-2:0]}}} \\
				\lstick{\ket{X_{ref}[n-1:0]}} &\qw & \gate{-} & \gate{\times} & \qw & \\
				\lstick{\ket{0}} & \targ & \qw & \ctrl{-1} & \qw & \rstick{\ket{x_{n-1}}} 
			}
			\]    
			
			\caption{Circuit of the horizontal shear: Step 2.}
			\label{h-shear-step-2} 	
		\end{subfigure} 
		\\
		\begin{subfigure}[tbhp]{3in}
			\[
			\Qcircuit @C = .4em @R = .7em @!R{
				\lstick{\ket{y_{[n-1:0]}}} & \qw & 	\qw & \qw & \qw & \qw & \rstick{\ket{y_{[n-1:0]}}} \\
				\lstick{\ket{\beta_{[n-1:0]}}} & \qw & \qw & \ctrl{3} & \qw &  \qw & \rstick{\ket{\beta_{[n-1:0]}}} \\
				\lstick{\ket{x_{n-1}}} &\ctrl{3} &  \ctrl{2} & \qw & \qw &  \qw & \rstick{\ket{x_{n-1}}} \\
				\lstick{\ket{x_{[n-2:0]}}} &\qw & \ctrl{1} & \qw & \qw &  \qw & \rstick{\ket{x_{[n-2:0]}}} \\
				\lstick{\ket{X_{ref}[n-1:0]}} &\qw & \gate{-} & \gate{\times} & \gate{IP} & \qw &  \\
				\lstick{\ket{0}} & \targ & \qw & \ctrl{-1} & \qw &  \qw & \rstick{\ket{x_{n-1}}} 
			}
			\]    
			
			\caption{Circuit of the horizontal shear: Step 3.}
			\label{h-shear-step-3} 
		\end{subfigure} 
		\\
		\begin{subfigure}[tbhp]{3in}
			\[
			\Qcircuit @C = .4em @R = .7em @!R{
				\lstick{\ket{y_{[n-1:0]}}} & \qw & 	\qw & \qw & \qw & \ctrl{4} & \qw & \rstick{\ket{y_{[n-1:0]}}} \\
				\lstick{\ket{\beta_{[n-1:0]}}} & \qw & \qw & \ctrl{3} & \qw &  \qw &  \qw & \rstick{\ket{\beta_{[n-1:0]}}} \\
				\lstick{\ket{x_{n-1}}} &\ctrl{3} &  \ctrl{2} & \qw & \qw &  \qw &  \qw & \rstick{\ket{x_{n-1}}} \\
				\lstick{\ket{x_{[n-2:0]}}} &\qw & \ctrl{1} & \qw & \qw &  \qw &  \qw & \rstick{\ket{x_{[n-2:0]}}} \\
				\lstick{\ket{X_{ref}[n-1:0]}} &\qw & \gate{-} & \gate{\times} & \gate{IP} & \gate{+} & \qw & \rstick{\ket{x_s}} \\
				\lstick{\ket{0}} & \targ & \qw & \ctrl{-1} & \qw &  \qw &  \qw & \rstick{\ket{x_{n-1}}} 
			}
			\]    
			
			\caption{Circuit of the horizontal shear: Step 4.  $x_s$ is the result of computation.}
			\label{h-shear-step-4} 
		\end{subfigure} 
		\caption{Quantum circuit for the horizontal shear of images for pixel located at values greater then the centroid value.  Adapted from \cite{FeiYan2017QuantumRotation}.}
		\label{top-h-part-image}
	\end{tcolorbox}
\end{figure}
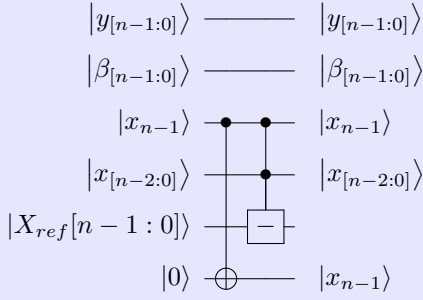
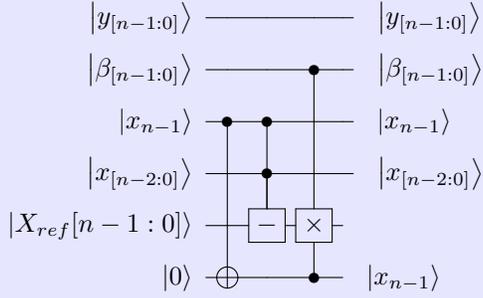
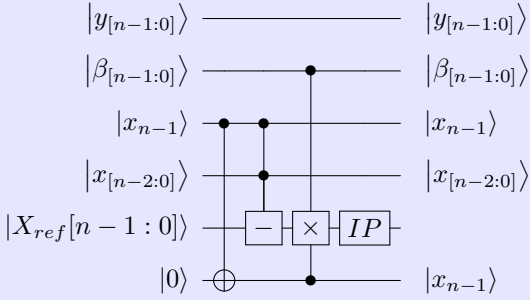
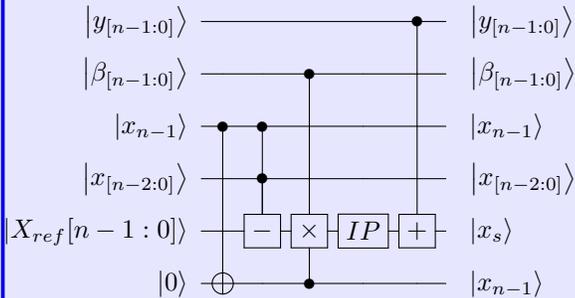

Figures \ref{top-h-part-image} illustrates the circuit to implement the horizontal shear for the case of images pixels located at values greater then the centroid value $X_{ref}$.  The circuit shown calculates equation \ref{encyclopedia-equation-p8} where $\alpha = tan \left(\frac{\theta}{2} \right)$.  Figure \ref{top-v-part-image} illustrates the circuits to implement the vertical shear for the case of images pixels located at values greater then the centroid value $Y_{ref}$.  The circuit shown calculates equation \ref{encyclopedia-equation-p9} and $\beta = sin(\theta)$.  The circuits take as input the pixel location values $(x,y)$.  The sinusoidal terms $\alpha = tan \left(\frac{\theta}{2} \right)$ for the horizontal shear and $sin(\theta) = \beta$ for the vertical shear are provided as inputs to the circuit.  The purpose of the interpolation module (labeled \textit{IP} in the Figures) is to round any fractional results from multiplication to the nearest whole number.   An interpolation circuit such as those shown in \cite{Edgard2018bilinear} \cite{WangLuo2015LNNInterpolation} can be used to implement this functional block.  

The horizontal shear circuit shown in Figure \ref{top-h-part-image} computes Equation \ref{encyclopedia-equation-p11} with four steps which are as follows.  Figure \ref{top-h-part-image} shows the quantum circuitry for the case of $x > X_{ref}$ where $X_{ref}$ is the centroid of rotation.

\begin{itemize}
	
	\item \textbf{Step 1:} If $x \leq X_{ref}$, use a quantum subtraction circuit compute $Y_{ref} - y$.  If $x > X_{ref}$, use a quantum subtraction circuit compute $y - Y_{ref}$.  
	
	\item \textbf{Step 2:} With a quantum controlled multiplication circuit, multiply the result of Step 1 by $tan \left( \frac{\theta}{2} \right)$  
	
	\item \textbf{Step 3:} With a quantum circuit implementation of the interpolation operation, round the result of Step 2 to the nearest integer
	
	\item \textbf{Step 4:} If $x \leq X_{ref}$, use a quantum subtraction circuit to subtract x by the result of Step 3.  If $x > X_{ref}$, use a quantum addition circuit to add x by the result of Step 3. 
	
\end{itemize}

Similarly, the vertical shear circuit shown in Figure \ref{top-v-part-image} computes Equation \ref{encyclopedia-equation-p10} with four steps which are as follows.  Figure \ref{top-v-part-image} shows the quantum circuitry for the case of $y > Y_{ref}$ where $Y_{ref}$ is the centroid of rotation.

\begin{itemize}
	
	\item \textbf{Step 1:} If $y \leq Y_{ref}$, use a quantum subtraction circuit compute $X_{ref} - x$.  If $y > Y_{ref}$, use a quantum subtraction circuit compute $x - X_{ref}$.
	
	\item \textbf{Step 2:} With a quantum controlled multiplication circuit, multiply the result of Step 2 by $sin \left( \frac{\theta}{2} \right)$  
	
	\item \textbf{Step 3:} With a quantum circuit implementation of the interpolation operation, round the result of Step 2 to the nearest integer
	
	\item \textbf{Step 4:} If $y \leq Y_{ref}$, use a quantum subtraction circuit to subtract y by the result of Step 3.  If $y > Y_{ref}$, use a quantum addition circuit to add y by the result of Step 3. 
	
\end{itemize}

\section{Conclusion}

This work introduces the design and evaluation of quantum circuits which can, in turn, be used to implement quantum algorithms.  We start with a brief introduction to quantum computation illustrating, at a high level, fundamental properties of quantum computers (superposition, entanglement, measurement and unitarity).  Then, our discussion introduces quantum operations (or gates) used to implement the quantum circuits shown in this work.  We focus on the Clifford+T gate set as there exists fault tolerant implementations for all these gates.  Our discussion of gates also illustrates how we can combine the Clifford+T gates to synthesize higher level logic gates.  After gates are introduced, the measures we use to evaluate the resource usage of quantum circuits is presented.  Fault tolerant quantum computation and circuits for NISQ machines have differing sets of measures and we introduce them.  Having seen examples of basic logic gates used in quantum computation, we now turn our attention to the design of quantum circuits for basic operations.  Arithmetic functions (addition, subtraction, multiplication and division) are chosen because many quantum algorithms make use of at least one of these arithmetic functions (see \cite{Bowregard} \cite{LeGall2017TriangleFinding} \cite{HuWenWen2019QuantumWatermarking} \cite{Bhaskar2016QuantumScientificComputation}).  For each circuit, we show the high level architecture and summarize how the logic gates shown previously are combined to form these arithmetic circuits.  We also describe how the arithmetic units perform their functions.   Afterward, we turn our attention to how we calculate the resource costs of quantum circuits such as quantum circuits for arithmetic.  We show the calculations of the resource cost measures for both NISQ and fault tolerant computers using quantum arithmetic circuits as examples.  We conclude by illustrating how building blocks (such as quantum arithmetic circuits) can be used to implement data paths for quantum algorithms.  We present the quantum circuit implementation of the image rotation operation.  The image rotation operation is a fundamental step for higher level image processing algorithms.  

The design of quantum computers is at an exciting stage.  The develop of new algorithms is an active area of research and shall continue to be so in the near future.  There already exist quantum computers of sufficient size to begin implementing quantum algorithms and circuits (albeit at a small scale).  Work continues to enlarge the size of the available quantum machines and improve the underlying technology.  New platforms (such as quantum dots or anyons \cite{Hasan2010QuantuTopologicalMachines} \cite{Petit2020SiQuantumUnit}) are being examined for their potential use in the development of quantum machines.  To be able to realize the performance gains promised by proposed quantum algorithms, quantum circuits and quantum data paths must be developed.  This article presents an introduction to the design and resource cost assessment of quantum circuits.  This article also presents an end-to-end implementation of a larger quantum data path system from basic gates to complete system.  We conclude that this article can provide a road map for scholars who wish to contribute to the field of quantum circuit design.

\bibliographystyle{IEEEtran}
\bibliography{encyclopedia-bibliography.bib}

\begin{thebibliography}{10}
\providecommand{\url}[1]{#1}
\csname url@samestyle\endcsname
\providecommand{\newblock}{\relax}
\providecommand{\bibinfo}[2]{#2}
\providecommand{\BIBentrySTDinterwordspacing}{\spaceskip=0pt\relax}
\providecommand{\BIBentryALTinterwordstretchfactor}{4}
\providecommand{\BIBentryALTinterwordspacing}{\spaceskip=\fontdimen2\font plus
\BIBentryALTinterwordstretchfactor\fontdimen3\font minus
  \fontdimen4\font\relax}
\providecommand{\BIBforeignlanguage}[2]{{%
\expandafter\ifx\csname l@#1\endcsname\relax
\typeout{** WARNING: IEEEtran.bst: No hyphenation pattern has been}%
\typeout{** loaded for the language `#1'. Using the pattern for}%
\typeout{** the default language instead.}%
\else
\language=\csname l@#1\endcsname
\fi
#2}}
\providecommand{\BIBdecl}{\relax}
\BIBdecl

\bibitem{munoz-coreas_everything_2022}
\BIBentryALTinterwordspacing
E.~Muñoz-Coreas and H.~Thapliyal, ``\BIBforeignlanguage{en}{Everything {You}
  {Always} {Wanted} to {Know} about {Quantum} {Circuits}},'' in
  \emph{\BIBforeignlanguage{en}{Wiley {Encyclopedia} of {Electrical} and
  {Electronics} {Engineering}}}.\hskip 1em plus 0.5em minus 0.4em\relax John
  Wiley \& Sons, Ltd, 2022, pp. 1--17. [Online]. Available:
  \url{https://onlinelibrary.wiley.com/doi/abs/10.1002/047134608X.W8440}
\BIBentrySTDinterwordspacing

\bibitem{Shor1997GeneralNonImageQuantumAlgorithm}
\BIBentryALTinterwordspacing
P.~W. Shor, ``\BIBforeignlanguage{eng}{Polynomial-time algorithms for prime
  factorization and discrete logarithms on a quantum computer},''
  \emph{\BIBforeignlanguage{eng}{SIAM Journal on Computing}}, vol.~26, no.~5,
  p. 1484, 1997. [Online]. Available:
  \url{http://search.proquest.com/docview/919434747/}
\BIBentrySTDinterwordspacing

\bibitem{Cheung}
D.~Cheung, D.~Maslov, J.~Mathew, and D.~K. Pradhan.\hskip 1em plus 0.5em minus
  0.4em\relax Berlin, Heidelberg: Springer Berlin Heidelberg, 2008, ch. On the
  Design and Optimization of a Quantum Polynomial-Time Attack on Elliptic Curve
  Cryptography, pp. 96--104.

\bibitem{Montanaro}
\BIBentryALTinterwordspacing
A.~Montanaro, ``Quantum pattern matching fast on average,''
  \emph{Algorithmica}, vol.~77, no.~1, pp. 16--39, Jan 2017. [Online].
  Available: \url{https://doi.org/10.1007/s00453-015-0060-4}
\BIBentrySTDinterwordspacing

\bibitem{Seroussi}
\BIBentryALTinterwordspacing
W.~{van Dam} and G.~{Seroussi}, ``Efficient quantum algorithms for estimating
  gauss sums,'' July 2002. [Online]. Available:
  \url{http://www.hpl.hp.com/techreports/2002/HPL-2002-208.pdf}
\BIBentrySTDinterwordspacing

\bibitem{VanDam2008exponential}
W.~van Dam and I.~E. Shparlinski, ``Classical and quantum algorithms for
  exponential congruences,'' in \emph{Theory of Quantum Computation,
  Communication, and Cryptography}, Y.~Kawano and M.~Mosca, Eds.\hskip 1em plus
  0.5em minus 0.4em\relax Berlin, Heidelberg: Springer Berlin Heidelberg, 2008,
  pp. 1--10.

\bibitem{LeGall2017TriangleFinding}
F.~Le~Gall and S.~Nakajima, ``\BIBforeignlanguage{{English}}{{Quantum Algorithm
  for Triangle Finding in Sparse Graphs}},''
  \emph{\BIBforeignlanguage{{English}}{Algorithmica}}, vol.~{79}, no.~{3}, pp.
  {941--959}, NOV {2017}, {26th International Symposium on Algorithms and
  Computation (ISAAC), Nagoya, Japan, Dec 09-11, 2015}.

\bibitem{Bowregard}
S.~Beauregard, ``{Circuit for Shor's algorithm using 2n+3 gubits},''
  \emph{Quantum Information \& Computation}, vol.~{3}, no.~{2}, pp. {175--185},
  {Mar} {2003}.

\bibitem{Aditya2018QuantumImageProcessing}
\BIBentryALTinterwordspacing
A.~{Dendukuri} and K.~{Luu}, ``{Image Processing in Quantum Computers},''
  \emph{arXiv e-prints}, p. arXiv:1812.11042, Dec. 2018. [Online]. Available:
  \url{https://arxiv.org/abs/1812.11042}
\BIBentrySTDinterwordspacing

\bibitem{HuWenWen2019QuantumWatermarking}
W.~Hu, R.-G. Zhou, and Y.~Li, ``\BIBforeignlanguage{{English}}{{Quantum
  Watermarking Based on Neighbor Mean Interpolation and LSB Steganography
  Algorithms}},'' \emph{\BIBforeignlanguage{{English}}{International Journal of
  Theoretical Physics}}, vol.~{58}, no.~{7}, pp. {2134--2157}, JUL {2019}.

\bibitem{ChengZhenwen2018QuantumWatermarking}
Z.~Qu, Z.~Cheng, M.~Wang, and W.~Liu, ``\BIBforeignlanguage{{English}}{{High
  Efficient and Robust Quantum Watermarking Algorithm Based on Quantum
  Log-polar Images and Matrix Coding}},''
  \emph{\BIBforeignlanguage{{English}}{Journal of Internet Technology}},
  vol.~{19}, no.~{6}, pp. {1831--1839}, NOV {2018}.

\bibitem{Bhaskar2016QuantumScientificComputation}
M.~K. Bhaskar, S.~Hadfield, A.~Papageorgiou, and I.~Petras,
  ``\BIBforeignlanguage{{English}}{Quantum algorithms and circuits for
  scientific computing},'' \emph{\BIBforeignlanguage{{English}}{Quantum
  Information \& Computation}}, vol.~{16}, no. {3-4}, pp. {197--236}, {Mar}
  {2016}.

\bibitem{Hallgren2007GeneralNonImageQuantumAlgorithm}
S.~Hallgren, ``\BIBforeignlanguage{eng}{Polynomial-time quantum algorithms for
  pell's equation and the principal ideal problem},''
  \emph{\BIBforeignlanguage{eng}{Journal of the ACM (JACM)}}, vol.~54, no.~1,
  pp. 1--19, 2007.

\bibitem{Bravyi2011QuantumImageAlgorithm}
S.~Bravyi, A.~W. Harrow, and A.~Hassidim, ``\BIBforeignlanguage{eng}{Quantum
  algorithms for testing properties of distributions},''
  \emph{\BIBforeignlanguage{eng}{IEEE Transactions on Information Theory}},
  vol.~57, no.~6, pp. 3971--3981, 2011.

\bibitem{Jordan2005science}
\BIBentryALTinterwordspacing
S.~P. Jordan, ``Fast quantum algorithm for numerical gradient estimation,''
  \emph{Phys. Rev. Lett.}, vol.~95, p. 050501, Jul 2005. [Online]. Available:
  \url{https://link.aps.org/doi/10.1103/PhysRevLett.95.050501}
\BIBentrySTDinterwordspacing

\bibitem{Trout2015QuantumScientificCompiletion}
\BIBentryALTinterwordspacing
C.~J. Trout and K.~R. Brown, ``Magic state distillation and gate compilation in
  quantum algorithms for quantum chemistry,'' \emph{International Journal of
  Quantum Chemistry}, vol. 115, no.~19, pp. 1296--1304, 2015. [Online].
  Available: \url{https://onlinelibrary.wiley.com/doi/abs/10.1002/qua.24856}
\BIBentrySTDinterwordspacing

\bibitem{Zoo}
S.~Jordan, \emph{Quantum Algorithm Zoo}, 2016, available at:
  https://quantumalgorithmzoo.org/.

\bibitem{IBM2017QuantumNISQMachine}
IBM, \emph{Quantum Computing - IBM Q}, 2021, available at:
  https://www.research.ibm.com/ibm-q/.

\bibitem{Hannes2017iontrapNISQ}
H.~Bernien, S.~Schwartz, A.~Keesling, H.~Levine, A.~Omran, H.~Pichler, S.~Choi,
  A.~S. Zibrov, M.~Endres, M.~Greiner, V.~Vuletić, and M.~D. Lukin, ``Probing
  many-body dynamics on a 51-atom quantum simulator,'' \emph{Nature}, vol. 551,
  no. 7682, pp. 579--584, 2017.

\bibitem{IonQ2020QuantumNISQMachine}
{IonQ, Incorporated }, \emph{IonQ | Trapped Ion Quantum Computing}, 2020,
  available at: https://ionq.com/.

\bibitem{Honeywell2020NISQmachine}
``\BIBforeignlanguage{English}{Honeywell introduces its next-generation quantum
  computer system model h1 honeywell has a cross-disciplinary team of over 150
  scientists, engineers, and software developers dedicated to advancing quantum
  computing},'' \emph{\BIBforeignlanguage{English}{Dataquest}}, Oct 30 2020,
  copyright - Copyright 2020 Cyber Media (India) Ltd., distributed by
  Contify.com;.

\bibitem{Edgard2019wow}
\BIBentryALTinterwordspacing
T.~S. {Humble}, H.~{Thapliyal}, E.~{Muñoz-Coreas}, F.~A. {Mohiyaddin}, and
  R.~S. {Bennink}, ``Quantum computing circuits and devices,'' \emph{IEEE
  Design \& Test}, vol.~36, no.~3, pp. 69--94, 2019. [Online]. Available:
  \url{https://ieeexplore.ieee.org/document/8681202}
\BIBentrySTDinterwordspacing

\bibitem{Maslov}
M.~Amy, D.~Maslov, M.~Mosca, and M.~Roetteler, ``A meet-in-the-middle algorithm
  for fast synthesis of depth-optimal quantum circuits,'' \emph{IEEE
  Transactions on Computer-Aided Design of Integrated Circuits and Systems},
  vol.~32, no.~6, pp. 818--830, June 2013.

\bibitem{Devitt}
S.~J. {Devitt}, A.~M. {Stephens}, W.~J. {Munro}, and K.~{Nemoto},
  ``{Requirements for fault-tolerant factoring on an atom-optics quantum
  computer},'' \emph{Nature Communications}, vol.~4, p. 2524, Oct. 2013.

\bibitem{Paler_2017FaultTolerant}
\BIBentryALTinterwordspacing
A.~Paler, I.~Polian, K.~Nemoto, and S.~J. Devitt, ``Fault-tolerant, high-level
  quantum circuits: form, compilation and description,'' \emph{Quantum Science
  and Technology}, vol.~2, no.~2, p. 025003, apr 2017. [Online]. Available:
  \url{https://doi.org/10.1088/2058-9565/aa66eb}
\BIBentrySTDinterwordspacing

\bibitem{Kiteav2005faulttolerant}
\BIBentryALTinterwordspacing
S.~Bravyi and A.~Kitaev, ``Universal quantum computation with ideal clifford
  gates and noisy ancillas,'' \emph{Phys. Rev. A}, vol.~71, p. 022316, Feb
  2005. [Online]. Available:
  \url{https://link.aps.org/doi/10.1103/PhysRevA.71.022316}
\BIBentrySTDinterwordspacing

\bibitem{Fowler2012QuantumSurfaceCodes}
A.~Fowler, M.~Mariantoni, J.~Martinis, and A.~Cleland,
  ``\BIBforeignlanguage{English}{Surface codes: Towards practical large-scale
  quantum computation},'' \emph{\BIBforeignlanguage{English}{Physical Review
  A}}, vol.~86, no.~3, 2012.

\bibitem{Steane1997SteaneCodes}
\BIBentryALTinterwordspacing
A.~M. Steane, ``Active stabilization, quantum computation, and quantum state
  synthesis,'' \emph{Phys. Rev. Lett.}, vol.~78, pp. 2252--2255, Mar 1997.
  [Online]. Available:
  \url{https://link.aps.org/doi/10.1103/PhysRevLett.78.2252}
\BIBentrySTDinterwordspacing

\bibitem{LaoLingling2021QuantumCircuitCOmpilation}
L.~Lao, H.~van Someren, I.~Ashraf, and C.~G. Almudever, ``Timing and
  resource-aware mapping of quantum circuits to superconducting processors,''
  \emph{IEEE Transactions on Computer-Aided Design of Integrated Circuits and
  Systems}, pp. 1--1, 2021.

\bibitem{Jayashree}
\BIBentryALTinterwordspacing
H.~V. Jayashree, H.~Thapliyal, H.~R. Arabnia, and V.~K. Agrawal,
  ``Ancilla-input and garbage-output optimized design of a reversible quantum
  integer multiplier,'' \emph{The Journal of Supercomputing}, vol.~72, no.~4,
  pp. 1477--1493, 2016. [Online]. Available:
  \url{http://dx.doi.org/10.1007/s11227-016-1676-0}
\BIBentrySTDinterwordspacing

\bibitem{Edgard2019multiplication}
\BIBentryALTinterwordspacing
E.~{Muñoz-Coreas} and H.~{Thapliyal}, ``Quantum circuit design of a {T-count}
  optimized integer multiplier,'' \emph{IEEE Transactions on Computers},
  vol.~68, no.~5, pp. 729--739, 2019. [Online]. Available:
  \url{https://ieeexplore.ieee.org/document/8543237}
\BIBentrySTDinterwordspacing

\bibitem{Kotiyal2014multiplier}
S.~Kotiyal, H.~Thapliyal, and N.~Ranganathan,
  ``\BIBforeignlanguage{eng}{Circuit for reversible quantum multiplier based on
  binary tree optimizing ancilla and garbage bits}.''\hskip 1em plus 0.5em
  minus 0.4em\relax IEEE, January 2014, pp. 545--550.

\bibitem{Edgard2021addition}
\BIBentryALTinterwordspacing
H.~Thapliyal, E.~Muñoz-Coreas, and V.~Khalus, ``Quantum circuit designs of
  carry lookahead adder optimized for {T-count} {T-depth} and qubits,''
  \emph{Sustainable Computing: Informatics and Systems}, vol.~29, p. 100457,
  2021. [Online]. Available:
  \url{https://www.sciencedirect.com/science/article/pii/S2210537920301815}
\BIBentrySTDinterwordspacing

\bibitem{Saeedi2}
\BIBentryALTinterwordspacing
I.~L. Markov and M.~Saeedi, ``Constant-optimized quantum circuits for modular
  multiplication and exponentiation,'' \emph{Quantum Information {\&}
  Computation}, vol.~12, no. 5-6, pp. 361--394, 2012. [Online]. Available:
  \url{http://www.rintonpress.com/xxqic12/qic-12-56/0361-0394.pdf}
\BIBentrySTDinterwordspacing

\bibitem{Lidia}
\BIBentryALTinterwordspacing
L.~Ruiz-Perez and J.~C. Garcia-Escartin, ``Quantum arithmetic with the quantum
  {Fourier} transform,'' \emph{Quantum Information Processing}, vol.~16, no.~6,
  p. 152, Apr 2017. [Online]. Available:
  \url{https://doi.org/10.1007/s11128-017-1603-1}
\BIBentrySTDinterwordspacing

\bibitem{Edgard2019divider}
H.~Thapliyal, E.~Mu\~{n}oz Coreas, T.~S.~S. Varun, and T.~S. Humble, ``Quantum
  circuit designs of integer division optimizing {T-count} and {T-depth},''
  \emph{IEEE Transactions on Emerging Topics in Computing}, vol.~9, no.~2, pp.
  1045--1056, April 2021.

\bibitem{Thapliyal2016addsub}
H.~Thapliyal, ``Mapping of subtractor and adder-subtractor circuits on
  reversible quantum gates,'' in \emph{Transactions on Computational Science
  XXVII}.\hskip 1em plus 0.5em minus 0.4em\relax Springer, 2016, pp. 10--34.

\bibitem{RiGuoZhou2017ImageInterpolation}
R.-G. Zhou, W.~Hu, P.~Fan, and H.~Ian,
  ``\BIBforeignlanguage{{English}}{{Quantum realization of the bilinear
  interpolation method for NEQR}},''
  \emph{\BIBforeignlanguage{{English}}{Scientific Reports}}, vol.~{7}, May
  {2017}.

\bibitem{FeiYan2017QuantumRotation}
F.~Yan, K.~Chen, S.~E. Venegas-Andraca, and J.~Zhao,
  ``\BIBforeignlanguage{{English}}{{Quantum image rotation by an arbitrary
  angle}},'' \emph{\BIBforeignlanguage{{English}}{{Quantum Information
  Processing}}}, vol.~{16}, no.~{11}, Nov {2017}.

\bibitem{Edgard2018bilinear}
\BIBentryALTinterwordspacing
E.~{Muñoz-Coreas} and H.~{Thapliyal}, ``{T-count} optimized quantum circuits
  for bilinear interpolation,'' in \emph{2018 Ninth International Green and
  Sustainable Computing Conference (IGSC)}, 2018, pp. 1--6. [Online].
  Available: \url{https://ieeexplore.ieee.org/document/8752146}
\BIBentrySTDinterwordspacing

\bibitem{Neto2020quantumNeuralNet}
F.~M. de~Paula~Neto, T.~B. Ludermir, W.~R. de~Oliveira, and A.~J. da~Silva,
  ``\BIBforeignlanguage{{English}}{{Implementing Any Nonlinear Quantum
  Neuron}},'' \emph{\BIBforeignlanguage{{English}}{{IEEE Transactions on Neural
  Networks and Learning Systems}}}, vol.~{31}, no.~{9}, pp. {3741--3746},
  {Sept} {2020}.

\bibitem{Edgard2018JETCsqrt}
\BIBentryALTinterwordspacing
E.~Mu\~{n}oz Coreas and H.~Thapliyal, ``T-count and qubit optimized quantum
  circuit design of the non-restoring square root algorithm,'' \emph{J. Emerg.
  Technol. Comput. Syst.}, vol.~14, no.~3, Oct. 2018. [Online]. Available:
  \url{https://doi-org.libproxy.library.unt.edu/10.1145/3264816}
\BIBentrySTDinterwordspacing

\bibitem{Zhang2020quantumMatrixOperations}
J.~Lin, D.-B. Zhang, S.~Zhang, T.~Li, X.~Wang, and W.-S. Bao,
  ``\BIBforeignlanguage{{English}}{{Quantum-enhanced least-square support
  vector machine: Simplified quantum algorithm and sparse solutions}},''
  \emph{\BIBforeignlanguage{{English}}{{Physics Letters A}}}, vol. {384},
  no.~{25}, SEP {2020}.

\bibitem{Quipper}
P.~Selinger{ et al.}, \emph{The Quipper System}, 2016, available at:
  http://www.mathstat.dal.ca/~selinger/quipper/doc/.

\bibitem{LIQUi}
D.~Wecker{ et al.}, \emph{Language-Integrated Quantum Operations:
  LIQUi$\vert\rangle$}, 2016, available at:
  https://www.microsoft.com/en-us/research/project/language-integrated-quantum-operations-liqui/.

\bibitem{Iliyasu2013useofrotation}
A.~M. Iliyasu, ``\BIBforeignlanguage{English}{Towards realising secure and
  efficient image and video processing applications on quantum computers},''
  \emph{\BIBforeignlanguage{English}{Entropy}}, vol.~15, no.~8, pp. 2874--2974,
  2013, copyright - Copyright MDPI AG 2013; Última actualización -
  2019-09-05.

\bibitem{Lehmann1999useofrotation}
T.~Lehmann, C.~Gonner, and K.~Spitzer, ``Survey: interpolation methods in
  medical image processing,'' \emph{IEEE Transactions on Medical Imaging},
  vol.~18, no.~11, pp. 1049--1075, 1999.

\bibitem{Haponen2003useofimagerotation}
A.~Happonen and U.~Ruotsalainen, ``Alignment of scans in dynamic {PET} study
  using stackgram domain,'' in \emph{2003 IEEE Nuclear Science Symposium.
  Conference Record (IEEE Cat. No.03CH37515)}, vol.~4, Oct 2003, pp. 2868--2872
  Vol.4.

\bibitem{Chaung2011QuantumBook}
M.~A. Nielsen and I.~L. Chuang, \emph{Quantum Computation and Quantum
  Information: 10th Anniversary Edition}, 10th~ed.\hskip 1em plus 0.5em minus
  0.4em\relax USA: Cambridge University Press, 2011.

\bibitem{sakurai_napolitano_2017}
J.~J. Sakurai and J.~Napolitano, \emph{Modern Quantum Mechanics}, 2nd~ed.\hskip
  1em plus 0.5em minus 0.4em\relax Cambridge University Press, 2017.

\bibitem{Rigetti2019QuantumNISQMachine}
{Rigetti Computing}, \emph{QPU Specifications - Rigetti 16Q Aspen 4}, 2019,
  available at: https://www.rigetti.com/qpu.

\bibitem{Zhou2000FaultTolerantQuantumComputation}
\BIBentryALTinterwordspacing
X.~Zhou, D.~W. Leung, and I.~L. Chuang, ``Methodology for quantum logic gate
  construction,'' \emph{Phys. Rev. A}, vol.~62, p. 052316, Oct 2000. [Online].
  Available: \url{https://link.aps.org/doi/10.1103/PhysRevA.62.052316}
\BIBentrySTDinterwordspacing

\bibitem{Griffiths2004Introduction}
D.~J. Griffiths, \emph{{Introduction to Quantum Mechanics (2nd Edition)}},
  2nd~ed.\hskip 1em plus 0.5em minus 0.4em\relax Pearson Prentice Hall, Apr.
  2004.

\bibitem{Haah2017magicstate}
\BIBentryALTinterwordspacing
J.~Haah, M.~B. Hastings, D.~Poulin, and D.~Wecker, ``Magic state distillation
  with low space overhead and optimal asymptotic input count,''
  \emph{{Quantum}}, vol.~1, p.~31, Oct. 2017. [Online]. Available:
  \url{https://doi.org/10.22331/q-2017-10-03-31}
\BIBentrySTDinterwordspacing

\bibitem{Kliuchnikov2013approximationofCliffT}
\BIBentryALTinterwordspacing
V.~Kliuchnikov, D.~Maslov, and M.~Mosca, ``Fast and efficient exact synthesis
  of single-qubit unitaries generated by clifford and t gates,'' \emph{Quantum
  Info. Comput.}, vol.~13, no. 7-8, pp. 607--630, Jul. 2013. [Online].
  Available: \url{http://dl.acm.org/citation.cfm?id=2535649.2535653}
\BIBentrySTDinterwordspacing

\bibitem{Miller}
D.~Miller, M.~Soeken, and R.~Drechsler, ``Mapping {NCV} circuits to optimized
  {Clifford+T} circuits,'' in \emph{Reversible Computation}, ser. Lecture Notes
  in Computer Science, S.~Yamashita and S.-i. Minato, Eds.\hskip 1em plus 0.5em
  minus 0.4em\relax Springer International Publishing, 2014, vol. 8507, pp.
  163--175.

\bibitem{Niel2000QFTalgorithm}
\BIBentryALTinterwordspacing
J.~{Niel de Beaudrap}, R.~{Cleve}, and J.~{Watrous}, ``{Sharp Quantum vs.
  Classical Query Complexity Separations},'' \emph{arXiv e-prints}, Nov. 2000.
  [Online]. Available: \url{https://arxiv.org/abs/quant-ph/0011065}
\BIBentrySTDinterwordspacing

\bibitem{Krovi2015QFTalgotithm}
\BIBentryALTinterwordspacing
H.~Krovi and A.~Russell, ``Quantum {Fourier} transforms and the complexity of
  link invariants for quantum doubles of finite groups,'' \emph{Communications
  in Mathematical Physics}, vol. 334, no.~2, pp. 743--777, 2015. [Online].
  Available: \url{https://doi.org/10.1007/s00220-014-2285-5}
\BIBentrySTDinterwordspacing

\bibitem{Takahashi}
Y.~Takahashi and N.~Kunihiro, ``A fast quantum circuit for addition with few
  qubits,'' \emph{Quantum Information \& Computation}, vol.~8, no. 6-7, p.
  0636–0649, 2008.

\bibitem{Bennett1973trashremoval}
\BIBentryALTinterwordspacing
C.~H. Bennett, ``Logical reversibility of computation,'' \emph{IBM J. Res.
  Dev.}, vol.~17, no.~6, pp. 525--532, Nov. 1973. [Online]. Available:
  \url{http://dx.doi.org/10.1147/rd.176.0525}
\BIBentrySTDinterwordspacing

\bibitem{Harty2014QuantumNISQErrorModels}
\BIBentryALTinterwordspacing
T.~P. Harty, D.~T.~C. Allcock, C.~J. Ballance, L.~Guidoni, H.~A. Janacek, N.~M.
  Linke, D.~N. Stacey, and D.~M. Lucas,
  ``\BIBforeignlanguage{eng}{High-fidelity preparation, gates, memory, and
  readout of a trapped-ion quantum bit.}''
  \emph{\BIBforeignlanguage{eng}{Physical review letters}}, vol. 113, no.~22,
  pp. 220\,501--220\,501, 2014. [Online]. Available:
  \url{http://search.proquest.com/docview/1826610841/}
\BIBentrySTDinterwordspacing

\bibitem{Alam2019QuantumNISQErrorModels}
\BIBentryALTinterwordspacing
M.~Alam, A.~Ash-Saki, and S.~Ghosh, ``Analysis of quantum approximate
  optimization algorithm under realistic noise in superconducting qubits,''
  \emph{arXiv e-prints}, 2019. [Online]. Available:
  \url{https://arxiv.org/abs/1907.09631}
\BIBentrySTDinterwordspacing

\bibitem{Cross2019NISQerror}
\BIBentryALTinterwordspacing
A.~W. Cross, L.~S. Bishop, S.~Sheldon, P.~D. Nation, and J.~M. Gambetta,
  ``Validating quantum computers using randomized model circuits,'' \emph{Phys.
  Rev. A}, vol. 100, p. 032328, Sep 2019. [Online]. Available:
  \url{https://link.aps.org/doi/10.1103/PhysRevA.100.032328}
\BIBentrySTDinterwordspacing

\bibitem{Kento2020NISQmetrics}
\BIBentryALTinterwordspacing
K.~{Oonishi}, T.~{Tanaka}, S.~{Uno}, T.~{Satoh}, R.~{Van Meter}, and
  N.~{Kunihiro}, ``{Efficient Construction of a Control Modular Adder on a
  Carry-Lookahead Adder Using Relative-phase Toffoli Gates},'' \emph{arXiv
  e-prints}, p. arXiv:2010.00255, Oct. 2020. [Online]. Available:
  \url{https://arxiv.org/abs/2010.00255}
\BIBentrySTDinterwordspacing

\bibitem{Paler2019NISQerror}
A.~{Paler}, D.~{Herr}, and S.~J. {Devitt}, ``Really small shoe boxes: On
  realistic quantum resource estimation,'' \emph{Computer}, vol.~52, no.~6, pp.
  27--37, June 2019.

\bibitem{oonishi2020addersqcla}
\BIBentryALTinterwordspacing
K.~Oonishi, T.~Tanaka, S.~Uno, T.~Satoh, R.~V. Meter, and N.~Kunihiro,
  ``Efficient construction of a control modular adder on a carry-lookahead
  adder using relative-phase toffoli gates,'' 2020. [Online]. Available:
  \url{https://arxiv.org/abs/2010.00255}
\BIBentrySTDinterwordspacing

\bibitem{Mosca2014FaultTolerantQuantumComputing}
M.~Amy, D.~Maslov, and M.~Mosca, ``Polynomial-time t-depth optimization of
  clifford+t circuits via matroid partitioning,'' \emph{IEEE Transactions on
  Computer-Aided Design of Integrated Circuits and Systems}, vol.~33, no.~10,
  pp. 1476--1489, Oct 2014.

\bibitem{HaiSheng2020FaultToleantImageCircuits}
H.-S. Li, P.~Fan, H.~Xia, H.~Peng, and G.~L. Long,
  ``\BIBforeignlanguage{{English}}{{Efficient quantum arithmetic operation
  circuits for quantum image processing}},''
  \emph{\BIBforeignlanguage{{English}}{{Science China-Physics Mechanics \&
  Astronomy}}}, vol.~{63}, no.~{8}, {JUN 4} {2020}.

\bibitem{PalerIOP}
\BIBentryALTinterwordspacing
A.~Paler, I.~Polian, K.~Nemoto, and S.~J. Devitt, ``Fault-tolerant, high-level
  quantum circuits: form, compilation and description,'' \emph{Quantum Science
  and Technology}, vol.~2, no.~2, p. 025003, 2017. [Online]. Available:
  \url{http://stacks.iop.org/2058-9565/2/i=2/a=025003}
\BIBentrySTDinterwordspacing

\bibitem{Fowler2008FaultTolerantQuantumComputing}
A.~Fowler, A.~M.~Stephens, and P.~Groszkowski, ``High threshold universal
  quantum computation on the surface code,'' \emph{Physical Review A}, vol.~80,
  03 2008.

\bibitem{Bombin}
\BIBentryALTinterwordspacing
H.~Bombin, ``Clifford gates by code deformation,'' \emph{New Journal of
  Physics}, vol.~13, no.~4, p. 043005, 2011. [Online]. Available:
  \url{http://stacks.iop.org/1367-2630/13/i=4/a=043005}
\BIBentrySTDinterwordspacing

\bibitem{Gosset}
\BIBentryALTinterwordspacing
D.~Gosset, V.~Kliuchnikov, M.~Mosca, and V.~Russo, ``An algorithm for the
  {T-count},'' \emph{Quantum Information {\&} Computation}, vol.~14, no. 15-16,
  pp. 1261--1276, 2014. [Online]. Available:
  \url{http://www.rintonpress.com/xxqic14/qic-14-1516/1261-1276.pdf}
\BIBentrySTDinterwordspacing

\bibitem{Kliuchnikov2016approximation}
V.~Kliuchnikov, D.~Maslov, and M.~Mosca, ``Practical approximation of
  single-qubit unitaries by single-qubit quantum clifford and t circuits,''
  \emph{IEEE Transactions on Computers}, vol.~65, no.~1, pp. 161--172, Jan
  2016.

\bibitem{Loke2017QFT}
\BIBentryALTinterwordspacing
S.~S. Zhou, T.~Loke, J.~A. Izaac, and J.~B. Wang, ``Quantum {Fourier} transform
  in computational basis,'' \emph{Quantum Information Processing}, vol.~16,
  no.~3, p.~82, Feb 2017. [Online]. Available:
  \url{https://doi.org/10.1007/s11128-017-1515-0}
\BIBentrySTDinterwordspacing

\bibitem{Fowler2004rotationgates}
\BIBentryALTinterwordspacing
A.~G. Fowler and L.~C.~L. Hollenberg, ``Scalability of {Shor's} algorithm with
  a limited set of rotation gates,'' \emph{Phys. Rev. A}, vol.~70, p. 032329,
  Sep 2004. [Online]. Available:
  \url{http://link.aps.org/doi/10.1103/PhysRevA.70.032329}
\BIBentrySTDinterwordspacing

\bibitem{Hales2000QFTalgorithm}
\BIBentryALTinterwordspacing
L.~Hales and S.~Hallgren, ``An improved quantum {Fourier} transform algorithm
  and applications,'' in \emph{Proceedings 41st Annual Symposium on Foundations
  of Computer Science}, 2000, pp. 515--525. [Online]. Available:
  \url{https://ieeexplore.ieee.org/document/892139}
\BIBentrySTDinterwordspacing

\bibitem{Soeken2017reciprocal}
M.~Soeken, M.~Roetteler, N.~Wiebe, and G.~D. Micheli, ``Design automation and
  design space exploration for quantum computers,'' in \emph{Design, Automation
  Test in Europe Conference Exhibition (DATE), 2017}, March 2017, pp. 470--475.

\bibitem{Cuccaro2004adder}
S.~A. {Cuccaro}, T.~G. {Draper}, S.~A. {Kutin}, and D.~{Petrie Moulton}, ``{A
  new quantum ripple-carry addition circuit},'' \emph{eprint
  arXiv:quant-ph/0410184}, Oct. 2004.

\bibitem{Lin}
\BIBentryALTinterwordspacing
C.-C. Lin, A.~Chakrabarti, and N.~K. Jha, ``Qlib: Quantum module library,''
  \emph{J. Emerg. Technol. Comput. Syst.}, vol.~11, no.~1, pp. 7:1--7:20, Oct.
  2014. [Online]. Available: \url{http://doi.acm.org/10.1145/2629430}
\BIBentrySTDinterwordspacing

\bibitem{Thapliyal2013adder}
\BIBentryALTinterwordspacing
H.~Thapliyal and N.~Ranganathan, ``Design of efficient reversible logic-based
  binary and bcd adder circuits,'' \emph{J. Emerg. Technol. Comput. Syst.},
  vol.~9, no.~3, pp. 17:1--17:31, Oct. 2013. [Online]. Available:
  \url{http://doi.acm.org/10.1145/2491682}
\BIBentrySTDinterwordspacing

\bibitem{Kowada}
\BIBentryALTinterwordspacing
L.~A.~B. Kowada, R.~Portugal, and C.~M.~H. de~Figueiredo, ``Reversible
  {Karatsuba's} algorithm,'' \emph{j-jucs}, vol.~12, no.~5, pp. 499--511, jun
  2006. [Online]. Available:
  \url{http://www.jucs.org/jucs\_12\_5/reversible\_karatsubas\_algorithm}
\BIBentrySTDinterwordspacing

\bibitem{Walter1993arraymultiplier}
C.~Walter, ``Systolic modular multiplication,'' \emph{IEEE Transactions on
  Computers}, vol.~42, no.~3, pp. 376--378, 1993.

\bibitem{Hwang1979Multiplier}
K.~Hwang, ``Global and modular two's complement cellular array multipliers,''
  \emph{IEEE Transactions on Computers}, vol. C-28, no.~4, pp. 300--306, 1979.

\bibitem{Aghababa2011QuantumDivisder}
A.~Khosropour, H.~Aghababa, and B.~Forouzandeh, ``Quantum division circuit
  based on restoring division algorithm,'' in \emph{2011 Eighth International
  Conference on Information Technology: New Generations}, April 2011, pp.
  1037--1040.

\bibitem{Lidia2017QFTcircuit}
\BIBentryALTinterwordspacing
L.~Ruiz-Perez, Garcia-Escartin, and J.~Carlos, ``Quantum arithmetic with the
  quantum {Fourier} transform,'' \emph{Quantum Information Processing},
  vol.~16, 2017. [Online]. Available:
  \url{https://doi.org/10.1007/s11128-017-1603-1}
\BIBentrySTDinterwordspacing

\bibitem{QFT-circuits}
\BIBentryALTinterwordspacing
A.~Pavlidis and E.~Floratos, ``{Quantum-{Fourier}-transform-based quantum
  arithmetic with qudits},'' \emph{Phys. Rev. A}, vol. 103, p. 032417, Mar
  2021. [Online]. Available:
  \url{https://link.aps.org/doi/10.1103/PhysRevA.103.032417}
\BIBentrySTDinterwordspacing

\bibitem{Selinger2013Tgatelayers}
\BIBentryALTinterwordspacing
P.~Selinger, ``Quantum circuits of {T-depth} one,'' \emph{Phys. Rev. A},
  vol.~87, p. 042302, Apr 2013. [Online]. Available:
  \url{https://link.aps.org/doi/10.1103/PhysRevA.87.042302}
\BIBentrySTDinterwordspacing

\bibitem{Niemann2019Tgatelayers}
P.~Niemann, A.~Gupta, and R.~Drechsler, ``{T-depth} optimization for
  fault-tolerant quantum circuits,'' in \emph{2019 IEEE 49th International
  Symposium on Multiple-Valued Logic (ISMVL)}, 2019, pp. 108--113.

\bibitem{YanFei2020CriticalQuantumImageProcessing}
\BIBentryALTinterwordspacing
F.~Yan, S.~E. Venegas-Andraca, and K.~Hirota, ``A critical and moving-forward
  view on quantum image processing,'' 2020. [Online]. Available:
  \url{https://arxiv.org/abs/2006.08747}
\BIBentrySTDinterwordspacing

\bibitem{Huifang2015useofrotation}
H.~Cheng and Y.~Wan, ``A new image rotation approach using radial basis
  functions,'' in \emph{2015 8th International Congress on Image and Signal
  Processing (CISP)}, 2015, pp. 1026--1031.

\bibitem{Yu2019QuantumImageTracking}
C.-H. Yu, F.~Gao, C.~Liu, D.~Huynh, M.~Reynolds, and J.~Wang,
  ``\BIBforeignlanguage{eng}{Quantum algorithm for visual tracking},''
  \emph{\BIBforeignlanguage{eng}{Physical Review A}}, vol.~99, no.~2, 2019.

\bibitem{Song2014QuantumImageEnceryption}
\BIBentryALTinterwordspacing
X.~{Song}, S.~{Wang}, J.~{Sang}, X.~{Yan}, and X.~{Niu}, ``Flexible quantum
  image secret sharing based on measurement and strip,'' in \emph{2014 Tenth
  International Conference on Intelligent Information Hiding and Multimedia
  Signal Processing}, 2014, pp. 215--218. [Online]. Available:
  \url{https://ieeexplore.ieee.org/document/6998306}
\BIBentrySTDinterwordspacing

\bibitem{Zhang2013quantumimagerepresentation}
\BIBentryALTinterwordspacing
Y.~Zhang, K.~Lu, Y.~Gao, and M.~Wang, ``{NEQR:} a novel enhanced quantum
  representation of digital images,'' \emph{Quantum Information Processing},
  vol.~12, pp. 2833 -- 2860, 2013. [Online]. Available:
  \url{https://doi.org/10.1007/s11128-013-0567-z}
\BIBentrySTDinterwordspacing

\bibitem{WangLuo2015LNNInterpolation}
N.~Jiang and L.~Wang, ``\BIBforeignlanguage{{English}}{{Quantum image scaling
  using nearest neighbor interpolation}},''
  \emph{\BIBforeignlanguage{{English}}{{Quantum Information Processing}}},
  vol.~{14}, no. {5, SI}, pp. {1559--1571}, {May} {2015}.

\bibitem{Hasan2010QuantuTopologicalMachines}
\BIBentryALTinterwordspacing
M.~Z. Hasan and C.~L. Kane, ``Colloquium: Topological insulators,'' \emph{Rev.
  Mod. Phys.}, vol.~82, pp. 3045--3067, Nov 2010. [Online]. Available:
  \url{https://link.aps.org/doi/10.1103/RevModPhys.82.3045}
\BIBentrySTDinterwordspacing

\bibitem{Petit2020SiQuantumUnit}
L.~Petit, H.~G.~J. Eenink, M.~Russ, W.~I.~L. Lawrie, N.~W. Hendrickx, S.~G.~J.
  Philips, J.~S. Clarke, L.~M.~K. Vandersypen, and M.~Veldhorst,
  ``\BIBforeignlanguage{English}{Universal quantum logic in hot silicon
  qubits},'' \emph{\BIBforeignlanguage{English}{Nature (London)}}, vol. 580,
  no. 7803, pp. 355--359, 2020.

\end{thebibliography}

\end{document}